\documentstyle[psfig]{mn}

\footnotesize
\newdimen\digitwidth    
\setbox0=\hbox{\rm0}
\digitwidth=\wd0
\catcode`!=\active
\def!{\kern\digitwidth}
\normalsize

\title{Glitches in Southern Pulsars}

\author[N. Wang et al.]{N. Wang,$^{1,2,3}$ R. N. Manchester,$^2$ 
R. T. Pace,$^2$\thanks{Present Address: Department of Physics and
Mathematical Physics, University of Adelaide, Adelaide SA 5005, Australia} 
M. Bailes,$^4$ V. M. Kaspi,$^{5,6}$ 
\newauthor B. W. Stappers$^7$\thanks{Present Address:
Astronomical Institute, University of Amsterdam, Kruislaan 403, 1098 SJ
Amsterdam, The Netherlands} and A. G. Lyne$^8$ \\
$^1$ Urumqi Astronomical Observatory, CAS, 40 South Beijing Road, Urumqi
830011, China; email:wangna@ms.xjb.ac.cn \\
$^2$ Australia Telescope National Facility, CSIRO, P.O. Box 76, Epping NSW 1710, 
Australia; email:rmanches@atnf.csiro.au \\
$^3$ Geophysics Department, Peking University, Beijing 100871, China\\
$^4$ Astrophysics and Supercomputing, Swinburne University of Technolgy, PO
Box 218, Hawthorn, Vic. 3122, Australia \\
$^5$ MIT, Center for Space Research 37-621, 70 Vassar St., Cambridge MA
02139, USA \\
$^6$ McGill University, Physics Department, 3600 University Street,
Montreal, Quebec, Canada H3A 2T8 \\
$^7$ Mt Stromlo Observatory, ANU, Private Bag, Weston Creek ACT 2611, Australia \\
$^8$ University of Manchester, Nuffield Radio Astronomy Laboratories,
Jodrell Bank, Macclesfield, Cheshire SK11 9DL, UK }

\begin{document}
\maketitle

\pagestyle{plain}
\begin{abstract}
Timing observations of 40 mostly young pulsars using the ATNF Parkes
radio telescope between 1990 January and 1998 December are
reported. In total, 20 previously unreported glitches and ten other
glitches were detected in 11 pulsars. These included 12 glitches in
PSR J1341$- $6220, corresponding to a glitch rate of 1.5 glitches per
year. We also detected the largest known glitch, in PSR J1614$-$5047, with
$\Delta\nu_g/\nu \approx 6.5 \times 10^{-6}$ where $\nu = 1/P$ is the
pulse frequency. Glitch parameters were determined both by
extrapolating timing solutions to inter-glitch intervals and by
phase-coherent timing fits across the glitch(es). These fits also gave
improved positions and dispersion measures for many of the
pulsars. Analysis of glitch parameters, both from this work and from
previously published results, shows that most glitches have a
fractional amplitude $\Delta\nu_g/\nu$ of between $10^{-8}$ and
$10^{-6}$. There is no consistent relationship between glitch
amplitude and the time since the previous glitch or the time to the
following glitch, either for the ensemble or for individual
pulsars. As previously recognised, the largest glitch activity is seen
in pulsars with ages of order 10$^4$ years, but for about 30 per cent
of such pulsars, no glitches were detected in the 8-year data
span. There is some evidence for a new type of timing irregularity in
which there is a significant increase in pulse frequency over a few
days, accompanied by a decrease in the magnitude of the slowdown
rate. Fits of an exponential recovery to post-glitch data show that
for most older pulsars, only a small fraction of the glitch decays. In
some younger pulsars, a large fraction of the glitch decays, but in
others, there is very little decay. Apart from the Crab pulsar, there
is no clear dependence of recovery timescale on pulsar age.
\end{abstract}

\begin{keywords}
stars: neutron -- pulsars: general
\end{keywords}

\section{Introduction}
Timing observations have shown that pulsars are remarkably stable clocks
(e.g. Kaspi, Taylor \& Ryba 1994)\nocite{ktr94}. However, they are not always
predictable. Two types of instability have been observed. In the first, the
pulsar period, $P$, or frequency $\nu = 1/P$, varies in an apparently random
fashion, fluctuating on timescales of days, weeks, months and years
\cite{cd85}. The intensity of the fluctuations can be described by an
`activity parameter' based on the amplitude of the second
frequency-derivative term in a Taylor-series fit to a set of pulse times of
arrival (TOAs):
\begin{equation}
\Delta(t) = \log\left(\frac{|\ddot\nu|}{6\nu} t^3\right)
\label{eq:activity}
\end{equation}
where $t$ is the length of the data span in
seconds \cite{antt94}. Conventionally $t=10^8$ s is adopted, giving the parameter
$\Delta_8$. In most young pulsars, the magnitude of $\ddot\nu$ is far in
excess of that expected from secular slowdown. Furthermore, timing noise is
normally very `red' \cite{cd85} and so the second derivative term is a good
indicator of the level of noise. Observations of large samples of pulsars
\cite{cd85,dsb+98} show that the activity parameter is positively
correlated with the value of $\dot P$, the first time-derivative of the
pulsar period.

In the second type of period instability, known as a `glitch', the pulse
frequency has a sudden increase which typically has a fractional amplitude
$\Delta\nu_g/\nu$ in the range $10^{-8}$ -- $10^{-6}$.  These glitches are
unpredictable but typically occur at intervals of a few years in young
pulsars. Coincident with the glitch, there is often an increase in the
magnitude of the frequency derivative, typically by $\sim 1$ per cent, which
sometimes decays with a timescale of weeks to years.  The
time-dependence of the pulsar frequency after a glitch is generally well
described by the following function:
\begin{equation}
\nu(t)=\nu_0(t)+\Delta\nu_g[1-Q(1- \exp(-t/\tau_d)] + \Delta\dot\nu_p t
\label{eq:glmodel}
\end{equation}
where $\nu_0(t)$ is the value of $\nu$ extrapolated from before the glitch,
$\Delta\nu_g = \Delta\nu_d + \Delta\nu_p$ is the total frequency change at the
time of the glitch ($t=0$), where $\Delta\nu_d$ is the part of the change
which decays exponentially and $\Delta\nu_p$ is the permanent change in
pulse frequency, $Q=\Delta\nu_d/\Delta\nu_g$, $\tau_d$ is the decay time constant and
$\Delta\dot\nu_p$ is the permanent change in $\dot\nu$ at the
time of the glitch. The increment in
frequency derivative is given by
\begin{equation}
\Delta\dot\nu(t)=\Delta\dot\nu_d\;\exp(-t/\tau_d) + \Delta\dot\nu_p =
\frac{-Q\Delta\nu_g}{\tau_d}\exp(-t/\tau_d) + \Delta\dot\nu_p.
\label{eq:glnudot}
\end{equation}

The middle term of equation \ref{eq:glmodel} was originally derived on the
basis of the two-component model for glitch recovery \cite{bppr69}, where
the initial frequency jump was due to a `starquake' or sudden change in
moment of inertia of the solid crust. In this model, the parameter $Q =
I_s/I_c$ where $I_s$ and $I_c$ are the moments of inertia of the superfluid
component and the crust, respectively.

In later versions of the theory, the frequency jump is due to a sudden
unpinning of vortex lines in crustal neutron superfluid 
\cite{ai75}. Post-glitch relaxation may be due to `vortex creep', that is, a
slow drift of the vortices across the crustal lattice (e.g. Alpar, Cheng \&
Pines 1989)\nocite{acp89} or drift of the crustal lattice itself with the
vortices remaining pinned \cite{rud91c,rzc98}. 

Depending on the internal temperature of the star
and other factors, the response to a glitch may be linear or
non-linear \cite{acp89}. In the linear regime, the frequency jump at $t=0$ decays
exponentially with time constant $\tau_d$. The fractional change in
frequency derivative at the jump is
\begin{equation}
\Delta\dot\nu_g/\dot\nu = (I_s/I)(\Delta\omega_s/\tau_d),
\end{equation}
where $\Delta\omega_s$ is the change at the time of the glitch in the
rotational lag of the superfluid, and $\omega_s = 2\pi(\nu_s - \nu)$, where
$\nu_s$ is the rotation frequency of the neutron superfluid. For glitches
which are large compared to the steady-state lag, $\Delta\omega_s \sim
-2\pi\Delta\nu$ and the change in spin-down rate can be large compared to
$I_s/I$.

In the non-linear regime, there is essentially no relaxation ($Q \approx 0$)
and
\begin{equation}
\Delta\dot\nu_g/\dot\nu = I_s/I,
\end{equation}
giving a permanent change $\Delta\dot\nu_p$ in $\dot\nu$
(Eqn~\ref{eq:glnudot}). Analyses of data for the Vela pulsar and other pulsars
\cite{acps88,accp93,rzc98} are generally consistent with a rather small
superfluid fraction, $I_s/I \sim 10^{-2}$, participating in the glitch
activity. In the Ruderman et al. (1998) model, permanent changes in
$\dot\nu$ may result from a change in the magnetic field configuration
associated with the crustal cracking at the time of a glitch.

Until the last few years, the number of known glitches and glitching pulsars
was modest. However, with relatively high-frequency searches at low Galactic
latitudes \cite{clj+92,jlm+92,kmj+92} discovering a much larger sample of
young pulsars, the number has increased significantly, with 21 pulsars
having 46 glitches being listed in the recent review by Lyne
(1996)\nocite{lyn96}. In this paper we describe timing observations
of a sample of 40 mostly young pulsars using the Parkes radio telescope which show a
total of 30 glitches in 11 pulsars.

\section{Observations and Analysis}
Observations were made using the Parkes 64-m radio telescope between
1990 January and 1998 December at frequencies around 430, 660, 1400
and 1650 MHz. The frequencies below 1 GHz were used only occasionally
to improve the determination of dispersion measures (DMs). At all
frequencies, cryogenic dual-channel receivers were used. Two separate
back-end systems were used to provide the frequency resolution
necessary for dedispersing. The early observations used filterbanks
and a one-bit digitizer system constructed at Jodrell Bank; a
$2\times256\times0.125$ MHz system was used at 430 MHz, a
$2\times128\times0.25$ MHz system was used at 660 MHz and at the
higher frequencies a $2\times64\times5$ MHz system was used. Further
details of the filterbank data acquisition system may be found in
Manchester et al. (1996)\nocite{mld+96}. From 1994 July, data were
acquired using a correlator system constructed at Caltech
\cite{nav94}. This system used two-bit digitization and an
autocorrelator with $2\times256$ lags over a maximum bandwidth of 128
MHz. From mid-1995, the lags were split between two frequency bands
which gave simultaneous observations at radio frequencies of 1400 and
1650 MHz \cite{sbm+97}.

For the filterbank systems, data were folded synchronously with the pulsar
period using off-line programs to give mean total intensity pulse
profiles. In the correlator system this folding was performed in a hardware
integrator having 1024 bins across the pulsar period. For both systems,
integration start times were established to better than $1 \mu$s using time
signals from the Observatory clock system. This was related to UTC(NIST)
using a radio link to the NASA DSN station at Tidbinbilla and clock offsets
kindly provided by the Jet Propulsion Laboratory, Pasadena. In subsequent
analysis, the data were summed to form 8 or 16 frequency subbands and time
subintegrations of 60 or 90 s duration and stored on disk. These files were
then summed in frequency and time using the best available period and DM
information to form a single profile for each observation. This was then
cross-correlated with a standard profile for the pulsar to give a pulse time
of arrival (TOA).

A total of 40 pulsars were monitored during the program. The J2000 and B1950
names of these pulsars, their periods and characteristic ages, and the dates
spanned by the observations are listed in Table~\ref{tb:obslist}.  Depending
on the strength of the pulsar, observation times for each TOA were between 2
and 12 min. Observations were obtained at intervals of between 2 and 6 weeks
for most of the pulsars in Table~\ref{tb:obslist}, with the known glitching
pulsars being observed more frequently. The TOAs resulting from this program
were analysed using version 11.3 of TEMPO\footnote{See
http://pulsar.princeton.edu/tempo/ for a description
of TEMPO.} which includes provision for fitting the glitch parameters given
in Equation~\ref{eq:glmodel}. The DE200 solar-system ephemeris \cite{sta90}
was used to convert TOAs to the solar-system barycentre. The sixth and
seventh columns of Table~\ref{tb:obslist} give the rms value of the TOA
uncertainty (dependent on the pulse period, pulse shape and signal/noise
ratio) and the rms residual after fitting for just pulse frequency and its
first derivative. For glitching pulsars, the largest glitch-free interval
was used for this fit. These residuals are dominated by the effects of timing
noise, and so they are an indication of its amplitude. An exception to this
is PSR J1513-5908, which has significant frequency second and third derivatives due
to secular slowdown \cite{kms+94}; in this case, the fit giving $\sigma_T$
included the first three frequency derivatives. 

\begin{table*}
\caption{Pulsars monitored for glitch activity}
\begin{tabular}{lcclcrrc}
PSR J  & PSR B & Period & Age & MJD Range  & $\sigma_{TOA}$ & $\sigma_T$ & Glitches \\
&&(s) & (ky) & &($\mu$s) &($\mu$s)&  detected\\
0536$-$7543 & 0538$-$75 & 1.2458 & 34937.0   & 48957--50827 & 396 &  7392  &  -- \\ 
0742$-$2822 & 0740$-$28 & 0.1668 & !!157.0   & 48932--51155 & 46  &  12777 &  -- \\ 
0835$-$4510 & 0833$-$45 & 0.0893 & !!!11.3   & 50024--51155 & 23  &  39668 &   1 \\ 
0908$-$4913 & 0906$-$49 & 0.1068 & !!111.6   & 48957--51155 & 57  &  25992 &  -- \\ 
0942$-$5552 & 0940$-$55 & 0.6644 & !!462.9   & 48874--51137 & 153 &  107551&  -- \\ \\
1001$-$5507 & 0959$-$54 & 1.4366 & !!442.9   & 47913--51137 & 125 &  7802  &  -- \\ 
1048$-$5832 & 1046$-$58 & 0.1237 & !!!20.4   & 47909--50940 & 49  &  99696 &  3 \\ 
1057$-$5226 & 1055$-$52 & 0.1971 & !!535.4   & 48814--51155 & 116 &  4569  &  -- \\ 
1105$-$6107 & --        & 0.0632 & !!!63.3   & 49176--51155 & 54  &  1176  &  2 \\ 
1123$-$6259 & --        & 0.2714 & !!818.6   & 49316--51155 & 288 &  1329  &  1 \\ \\
1133$-$6250 & 1131$-$62 & 1.0229 & 35855.0   & 48928--51094 & 815 &  5814  &  -- \\ 
1224$-$6407 & 1221$-$63 & 0.2165 & !!692.2   & 47912--51155 & 39  &  1309  &  -- \\
1320$-$5359 & 1317$-$53 & 0.2797 & !!478.9   & 50536--51155 & 181 &  434   &  -- \\
1326$-$5859 & 1323$-$58 & 0.4780 & !2358.5   & 50242--51094 & 72  &  2918  &  -- \\ 
1328$-$4357 & 1325$-$43 & 0.5327 & !2796.0   & 50738--51155 & 190 &  228   &  -- \\ \\
1341$-$6220 & --        & 0.1933 & !!!12.1   & 47915--51022 & 357 &  2309  &  12  \\ 
1359$-$6038 & 1356$-$60 & 0.1275 & !!318.7   & 50329--51155 & 18  &  102   &  -- \\
1435$-$5954 & --        & 0.4730 & !4857.5   & 49955--51135 & 413 &  2028  &  -- \\ 
1453$-$6413 & 1449$-$64 & 0.1795 & !1035.0   & 50669--51093 & 13  &  87    &  -- \\ \\
1456$-$6843 & 1451$-$68 & 0.2634 & 42244.9   & 48330--51094 & 96  &  1175  &  -- \\
1513$-$5908 & 1509$-$58 & 0.1502 & !!!!1.5   & 48296--51155 & 241 &  2485  &  -- \\     
1539$-$5626 & 1535$-$56 & 0.2434 & !!795.2   & 48874--51155 & 116 &  10560 &  -- \\
1549$-$4848 & --        & 0.2883 & !!323.8   & 49574--51045 & 299 &  1749  &  -- \\
1559$-$4438 & 1556$-$44 & 0.2571 & !3994.7   & 47910--51155 & 47  &  1728  &  -- \\
1559$-$5545 & 1555$-$55 & 0.9572 & !!740.5   & 49559--51135 & 277 &  36457 &  -- \\ \\
1602$-$5100 & 1558$-$50 & 0.8642 & !!196.3   & 48297--51155 & 139 &  207783&  -- \\
1614$-$5047 & 1610$-$50 & 0.2316 & !!!!7.4   & 48295--50926 & 148 &  19381 &  1 \\
1637$-$4553 & 1634$-$45 & 0.1188 & !!590.3   & 50669--51155 & 74  &  159   &  -- \\
1644$-$4559 & 1641$-$45 & 0.4551 & !!358.8   & 48956--51156 & 411 &  13055 &  --\\
1646$-$4346 & 1643$-$43 & 0.2316 & !!!32.5   & 47912--50502 & 350 &  162568&-- \\  \\
1709$-$4428 & 1706$-$44 & 0.1024 & !!!17.4   & 47909--51156 & 43  &  4862  &  1 \\ 
1722$-$3712 & 1719$-$37 & 0.2362 & !!345.9   & 49078--51094 & 58  &  10878 &  -- \\
1730$-$3350 & 1727$-$33 & 0.1394 & !!!26.0   & 50538--51155 & 96  &  10344 &  -- \\
1731$-$4744 & 1727$-$47 & 0.8297 & !!!80.3   & 49043--51156 & 62  &  8966  &  2 \\
1739$-$2903 & 1736$-$29 & 0.3229 & !!651.7   & 50739--51155 & 126 &  265   &  -- \\ \\  
1752$-$2806 & 1749$-$28 & 0.5626 & !!109.5   & 48145--51138 & 49  &  15282 &  -- \\
1801$-$2304 & 1758$-$23 & 0.4158 & !!!58.3   & 48296--51156 & 561 &  5140  &  4 \\
1801$-$2451 & 1757$-$24 & 0.1249 & !!!15.4   & 48896--50884 & 100 &  40947 &  2 \\ 
1803$-$2137 & 1800$-$21 & 0.1336 & !!!15.8   & 50669--51155 & 123 &  1603  &  1 \\
1822$-$4209 & --        & 0.4565 & 15770.5   & 49540--51138 & 431 &  1801  &  -- \\ 
\end{tabular}     
\label{tb:obslist}
\end{table*}

For several pulsars, the position and DM were measured with comparable or
better precision than was previously available. Positions were obtained
using the so-called `pre-whitening' method (Kaspi et al. 1994). In this
method, the position is fitted along with sufficiently many frequency-derivative
terms and in some cases, glitch terms, to `absorb' the timing noise and give
an approximately Gaussian distribution of timing residuals. DMs were
measured from short sections of data where there were multi-frequency
observations. This ensures that contamination by long-term timing noise is
not a problem and that error estimates are realistic.

For glitches where the amplitude of the glitch and the interval between the
last pre-glitch observation and the first post-glitch observation are not
too large, the epoch of the glitch can be determined by requiring that the
pulse phase be continuous over the glitch. In this situation, TEMPO gives an
estimate of the glitch parameters, including the epoch, and their
uncertainty. Where post-glitch decay parameters are estimated, these refer
to the long-term decay, typically with timescales of hundreds of days. The
resolution of our observations is generally not sufficient to detect the
more rapid post-glitch recoveries observed in some pulsars, for example, the
Vela pulsar \cite{fla90,mhmk90}.

For larger glitches where there may be one or more turns of phase in the
residuals between the bounding observations, the glitch epoch is
uncertain. Other glitch parameters are affected by this uncertainty. To
estimate uncertainties in this case, the following procedure was adopted. The
glitch epoch was taken to be halfway between the bounding observations, with
an uncertainty of half their separation. The increments $\Delta\nu_g$ and
$\Delta\dot\nu_g$ were then computed for an assumed glitch epoch $t_g$ by
extrapolating the pre- and post-glitch fits to $t_g$ and taking
differences. Uncertainties were similarly extrapolated and quadrature sums taken
for the uncertainties in the increments. This was done separately for $t_g$ at the
two bounding epochs. The final error estimates were then the quadrature sum
of the difference between the increments at either end of the data gap and
the larger of the uncertainties in the increments.

\section{Results}
Table~\ref{tb:glpsrs} lists parameters for the 11 pulsars for which glitches
were detected. The pulsar J2000 positions and DMs used in or derived from
the analysis are given. Uncertainties in the last digit quoted are given in
parentheses. These and other quoted uncertainties from TEMPO fits are twice
the formal rms error. Position fits were generally to the largest
available data span not too strongly affected by post-glitch post-glitch
recovery. As described above, the data spans used to determine the positions
and their errors were `pre-whitened' to eliminate the effects of period
noise from the positions and their errors. The fifth and sixth columns give
the data span used when fitting the position and the final rms timing
residual. References for the position and DM are given in the
final column.

\begin{table*}
\begin{minipage}{150mm}
\caption{Parameters for glitching pulsars}
\begin{tabular}{llllcrr}
  ~~~PSR J & ~R.A.(J2000)& ~Dec.(J2000) &  \multicolumn{1}{c}{DM} & MJD
  Range & $\sigma_W$ & Ref.
\footnote{Reference: 1. Taylor, Manchester \& Lyne (1993) \nocite{tml93};
2. Kaspi et al. (1996) \nocite{kbm+96}; 3. D'Amico et al. (1997)
\nocite{dsb+98}; 4. Frail, Kulkarni \& Vasisht (1993) \nocite{fkv93}; 5. This paper.}\\
 & ~(h~~m~~s) & ~~(\degr~~\arcmin~~\arcsec) & (cm$^{-3}$ pc) &  & ($\mu$s) & Posn,DM \\ \\

0835$-$4510 & 08:35:20.68(2)\footnote{Position is for epoch MJD 41380 with proper motion 
$\mu_{\alpha}$ = $-48$ mas yr$^{-1}$, $\mu_{\delta}$ = 35 mas yr$^{-1}$}
                              & $-$45:10:35.8(3)$^b$ & !!68.094(4) & $-$ & $-$ & 1,5 \\
1048$-$5832 & 10:48:12.2(1)   & $-$58:32:05.8(8) & !129.10(1)  & 49043--50536 & 1150 & 5,1 \\
1105$-$6107 & 11:05:26.17(4)  & $-$61:07:51.4(3) & !271.01(4)  & 49176--50402 & 1080 & 5,2 \\
1123$-$6259 & 11:23:55.549(12)& $-$62:59:10.74(9)& !223.26(3)  & 49708--51155 & 785  & 5,3 \\
1341$-$6220 & 13:41:42.63(8)  & $-$62:20:20.7(5) & !717.3(6)   & 48874--49888 & $-$  & 5,5 \\ 
1614$-$5047 & 16:14:11.29(3)  & $-$50:48:03.5(5) & !582.8(3)   & 50269--50778 & 736  & 5,1 \\
1709$-$4428 & 17:09:42.728(2) & $-$44:29:08.24(6)& !!75.69(5)  & 48928--51156 & 211  & 5,1 \\
1731$-$4744 & 17:31:42.103(5) & $-$47:44:34.56(14) & !123.33(2) &49415--50704 & 351  & 5,5 \\
1801$-$2304 & 18:01:19.829(9) & $-$23:04:44.2(2) & 1074(6)     & $-$ & $-$ & 4,1 \\
1801$-$2451 & 18:01:00.223(7) & $-$24:51:27.1(1) & !289(1)     & $-$ & $-$ & 1,1 \\
1803$-$2137 & 18:03:51.35(3)  & $-$21:37:07.2(5) & !233.9(3)   & $-$ & $-$ & 1,1 \\
\end{tabular}
\label{tb:glpsrs}
\end{minipage}
\end{table*}

A total of 30 glitches were observed in these 11 pulsars. Independent fits
to the pre-glitch, inter-glitch and post-glitch timing data are given in
Table~\ref{tb:pparam}. Except for short sections of data, the fits include
a $\ddot\nu$ term. In every case, this term is dominated by recovery from a
previous glitch; the $\ddot\nu$ from the long-term secular slow down is
negligible by comparison. Higher-order frequency derivative terms were
not fitted, so the rms residuals given in right-most column reflect
the presence of random period irregularities, especially for longer
data spans.

Glitch parameters are listed in Table~\ref{tb:glparam}. If
marked by $*$, the glitch epoch is determined by requiring phase continuity
across the glitch. The next two columns give the fractional steps in $\nu$ and
$\dot\nu$ at the glitch determined by extrapolating the pre- and post-glitch
solutions (Table~\ref{tb:pparam}) to the glitch epoch, with uncertainties
determined as described in Section 2. Parameters in the remaining columns
were determined using TEMPO to fit phases across the glitch.

Ten of these glitches have been previously reported: single glitches in PSR
J0835$-$4510 \cite{fla96}, PSR J1709$-$4428 \cite{jml+95,sl96}, PSR J1731$-$4744
\cite{dm97} and PSR J1801$-$2451 \cite{lkb+96}, and three glitches in PSR
J1341$-$6220 \cite{kmj+92,sl96} and PSR J1801$-$2306 \cite{klm+93,sl96}. We
reanalyse these glitches for completeness and consistency with the results
for the other pulsars.

In the following sections we discuss each pulsar in turn.

\begin{table*}
\caption{Pulse frequency parameters for glitching pulsars}
\begin{tabular}{lcllrccrc}
~~~PSR J & Int. & \multicolumn{1}{c}{$\nu$} & \multicolumn{1}{c}{$\dot\nu$} & 
       \multicolumn{1}{c}{$\ddot\nu$} & Epoch & Span & No. of & Residual \\ 
 & & \multicolumn{1}{c}{(s$^{-1}$)} & \multicolumn{1}{c}{($10^{-12}$ s$^{-2}$)} & 
       \multicolumn{1}{c}{($10^{-24}$ s$^{-3}$)} & (MJD) & (MJD) & TOAs & ($\mu$s)\\ \\
0835$-$4510 & ~~~~$-$1~~  & 11.1964839822(3)  & $-$15.58140(4)  & !!!873(10)   & 50155.00 & 50024--50364 & 38  & !!240 \\
            & ~~1$-$~~~~ & 11.1955738808(10) & $-$15.61780(6)  & !!!985(12)   & 50847.00 & 50387--51155 & 223 & !3190 \\ \\     
1048$-$5832 & ~~~~$-$1~~  & 8.087302320(3)    &  $-$6.27373(7)  & !!!!72(10)   & 48419.00 & 47909--48929 & 59 &  10120  \\ 
            &~~1$-$2~~  & 8.0869925648(4)   &  $-$6.2719(4)   & !!!!!--          & 48991.00 & 48957--49026 & 9 & !!113 \\ 
            & ~~2$-$3~~  & 8.0865824929(8)   &  $-$6.28460(3)  & !!!146.7(14) & 49790.00 & 49043--50786 & 164 &  21400 \\ 
            & ~~3$-$~~~~  & 8.08599240181(15) &  $-$6.29765(7)  & !!!390(40)   & 50889.00 & 50791--50940 & 51 & !!190 \\ \\
1105$-$6107 & ~~~~$-$1~~  & 15.8248916396(18) & $-$3.95819(5)  & !!$-$49(5)   & 49789.00 & 49176--50402 & 127 &  !6426 \\
            &~~1$-$2~~  & 15.8246473365(3)  & $-$3.96313(15) & !!!!!--         & 50516.00 & 50433--50599 & 27 &!!172 \\
            & ~~2$-$~~~~  & 15.8245521708(20) & $-$3.96325(10) & !!$-$54(18)  & 50794.00 & 50433--51155 & 94 & !4306  \\ \\
1123$-$6259 & ~~~~$-$1~~  & 3.68413905495(20) & $-$0.07125(5)  &  !!!!!--        & 49510.00 & 49316--49704 & 15  & !!998  \\
            & ~~1$-$~~~~  & 3.68413613509(3)  & $-$0.0712993(10) &  !!!!!--      & 50432.00 & 49708--51155 & 195 & !!817    \\ \\
1341$-$6220 & ~~~~$-$1~~  & 5.173922035(4)    & $-$6.7727      & !!!!!--         & 47942.00 & 47915--47969 & 9  & !1402 \\ 
            & ~~1$-$2~~  & 5.1737636204(9)   & $-$6.77427(7)  & !!$-$26(19)     & 48226.00 & 48011--48442 & 30 & !1778 \\
            & ~~2$-$3~~  & 5.1735752982(4)   & $-$6.7714(3)   & !!!!!--         & 48548.00 & 48465--48631 &16  & !!676 \\
            & ~~3$-$4~~  & 5.1733890567(14)  & $-$6.77215(7) & !!!190(30)       & 48875.00 & 48635--49114  & 26&!2309   \\ 
            & ~~4$-$5~~  & 5.1732030931(11)  & $-$6.7716(11)  & !!!!!--        & 49193.00 & 49159--49227 & 8 & !!418   \\
            & ~~5$-$6~~  & 5.173030116(14)   & $-$6.767       & !!!!!--        & 49490.00 & 49488--49491 & 2 & !!!!--  \\
            & ~~6$-$7~~  & 5.1729425780(5)   & $-$6.7677(3)   & !!!!!--        & 49640.00 & 49540--49739 & 9 &!!678 \\
            & ~~7$-$8~~  & 5.1728338894(5)   & $-$6.7659(3)   & !!!!!--        & 49826.00 & 49762--49889 & 13& !!411  \\ 
            & ~~8$-$9~~  & 5.172758037(3)    & $-$6.753(3)    & !!!!!--        & 49956.00 & 49920--49993 & 12& !1385   \\ 
            & ~~9$-$10  & 5.1726390106(3)   & $-$6.77150(3)  & !!!270(12)    & 50174.00 & 50025--50323 & 31 & !!395   \\
            & 10$-$11  & 5.1724946720(3)   & $-$6.77004(6)  & !!!240(50)    & 50421.00  & 50341--50501 & 19& !!191 \\ 
            & 11$-$12  & 5.1723883167(8)   & $-$6.76847(19) & !!!830(180)   & 50603.00 & 50536--50671 & 29 & !!482 \\
            & 12$-$~~~~  & 5.1722422270(5)   & $-$6.77115(7)  & !!$-$10(19)    & 50859.00 & 50696--51022 & 83 & !1102 \\ \\  
           
1614$-$5047 &  ~~~~$-$1~~  & 4.317954008(15)   & $-$9.1886(3)  & !$-$523(143)    & 48409.00 & 48295--48523 & 32 & !2054 \\
            &  ~~~~$-$1~~  & 4.3177762511(8)   & $-$9.1652(4)  & $-$1905(670)   & 48633.00 & 48596--48669 & 11 &  !!172 \\
            &  ~~~~$-$1~~  & 4.3175461311(4)   & $-$9.19054(4) & !!$-$73(12)      & 48923.00 & 48732--49114 & 23 & !!820 \\
	    &  ~~~~$-$1~~  & 4.317231719(4)    & $-$9.1456(7)  & !$-$211(227)    & 49320.00 & 49159--49482 & 18 & !5104 \\
            &  ~~~~$-$1~~  & 4.3169526367(20)  & $-$9.1552(4)   & !!!280(180)  & 49673.00 & 49559--49787 & 22& !2549  \\
            & ~~1$-$~~~~  & 4.3167356028(6)   & $-$9.23644(4)  & !!!350(30)  & 49981.00 & 49818--50143 & 52 & !1001  \\ 
            & ~~1$-$~~~~  & 4.3163251092(3)   & $-$9.225923(13)&  !$-$204(3)  & 50496.00 & 50214--50777 & 87 & !1239  \\ 
            & ~~1$-$~~~~  & 4.3160405830(14)  & $-$9.2199(3)   & $-$1600(300)& 50853.00 & 50780--50926 & 34 & !1132  \\ \\
1709$-$4428 & ~~~~$-$1~~  & 9.7612713776(3)   & $-$8.863749(8) & !!!124.1(14) & 48328.00 & 47909--48746 & 44& !!559 \\   
            & ~~1$-$~~~~  & 9.7608645977(8)   & $-$8.89001(13) & !!!181(14)   & 48885.00 & 48812--48959 & 21  & !!183  \\ 
            & ~~1$-$~~~~  & 9.7599781350(7)   & $-$8.857444(12)& !!!173.1(7)  & 50042.00 & 49000--51156 & 238 & !9857  \\ \\
1731$-$4744 &  ~~~~$-$1~~  & 1.20510332177(3)  & $-$0.237549(7) & !!!!!--        & 49204.00 & 49043--49364 & 24& !!375  \\
            & ~~1$-$2~~  & 1.20508592522(8) & $-$0.237668(3)  & !!!!!2.5(3)    & 50059.00  & 49415--50703  & 94 &!3403  \\ 
            & ~~2$-$~~~~  & 1.20506786194(6) & $-$0.237616(5)  & !!!!!5.6(13)   & 50939.00  & 50722--51156  & 47& !!565  \\ \\
1801$-$2304 & ~~~~$-$1~~  & 2.4051602072(9)  & $-$0.6527(6)     & !!!!!--         & 48357.00  & 48296--48417  & 11& !1451  \\
            & ~~1$-$2~~  & 2.40512168588(17)  & $-$0.653540(5)   & !!!!!2.7(6)    & 49054.00  & 48464--49644  & 58 & !2961  \\
            & ~~2$-$3~~  & 2.40507531889(14)  & $-$0.65347(5)    & !!!!!--         & 49878.00  & 49730--50027  & 27 &  !1107  \\
            & ~~3$-$4~~  & 2.40505493478(18)  & $-$0.65346(7)    & !!!!!--         & 50240.00  & 50116--50364  & 23 & !1018  \\
            & ~~4$-$~~~~  & 2.4050229993(3)    & $-$0.652923(12)  & !!!!47(3)       & 50809.00  & 50461--51156  & 100 &  !3246  \\ \\ 
1801$-$2451 & ~~~~$-$1~~  & 8.007176212(3)   & $-$8.17639(11)  &  !!!130(40)  & 49141.00  & 48896--49387  & 21& !2256 \\
            & ~~1$-$2~~  & 8.0065388371(19)  & $-$8.19045(7)   &  !!!399(7)    & 50064.00  & 49481--50647  & 89 & 12262  \\
            & ~~2$-$~~~~  & 8.0060495257(7)   & $-$8.20337(10)  &  !$-$140(60)  & 50770.00  & 50656--50884  & 64 & !!669  \\ \\

1803$-$2137 & ~~~~$-$1~~  & 7.4832982197(3)   & $-$7.4889(3)    &  !!!!!--      & 50710.00  & 50669--50751  & 16& !!176\\
            & ~~1$-$2~~  & 7.4832434845(19)  & $-$7.5440(6)    &  !!4500(700)  & 50831.00  & 50779--50883  & 13 & !!588 \\
            & ~~2$-$~~~~  & 7.4831211339(4)   & $-$7.52780(5)   &  !!!484(190)     & 51019.00  & 50883--51156  & 10 & !!152\\ \\
 
\end{tabular}   
\label{tb:pparam}
\end{table*}

\begin{table*}
\begin{minipage}{165mm}
\caption{Glitch parameters for eleven pulsars}
\begin{tabular}{lclllllllll}
 & & & &\multicolumn{2}{c}{Extrapolated} & \multicolumn{5}{c}{Fitted} \\
~~~PSR J & Glt. & \multicolumn{2}{c}{Glitch Epoch~~~~~} & $\Delta\nu_g/\nu$ &
$\Delta\dot\nu_g/\dot\nu$ & $\Delta\nu_g/\nu$ &  $\Delta\dot\nu_g/\dot\nu$ &
   \multicolumn{1}{c}{$Q$} & \multicolumn{1}{c}{$\tau_d$} & Resid. \\                   
& No.& (Date) & ~~(MJD)  & ($10^{-9}$) & ($10^{-3}$) & ($10^{-9}$) & ($10^{-3}$) & 
& \multicolumn{1}{c}{(d)} & ($\mu$s) \\ \\

0835$-$4510 & 1 & 961013 &50369.345(2)$^*$&  2110(17)   & !5.95(3)  & 2132(52) & --  & 0.38(2)  & 916(48) & 242 \\ \\
1048$-$5832 & 1 & 921118 & 48944(2)$^*$   &  !!25(2)    & !0.3(1)   & !!19(2)  & --  & --       & --      & 865 \\ 
            & 2 & 930216 & 49034(9)       &  2995(7)    & !3.7(1)   & 3000(10) & --  & 0.025(3) & 100     & 1630  \\
            & 3 & 971206 & 50788(3)       &  !771(2)    & !4.62(6)  & !769(3)   & --  & 0.245(3) & 400     & 465  \\ \\
1105$-$6107 & 1 & 961130 & 50417(16)      &  !281(3)    & !1.3(2.9) & !279.7(2) & 4.63(4) &-- &--  & 480  \\
            & 2 & 970613 & 50610(3)$^*$   &  !!!1.3(2)  & !0.19(1)  & !!!2.1(3) & --      &-- &--  & 420 \\ \\
1123$-$6259 & 1 & 941219 & 49705.87(1)$^*$ & !749.12(12) & !1.0(4)& !749.31(14)& -- & 0.0026(1)& 840(100) & 848  \\ \\
1341$-$6220 & 1 & 900408 & 47989(21)      &  1507(1)    & !0.15(6)  & 1509(1)   & -- & -- & -- & 2352 \\
            & 2 & 910716 & 48453(12)      &  !!24.2(9)  & !0.50(7)  & !!22.5(7) & $-$0.51(5) & -- & -- & -- \\
            & 3 & 920124 & 48645(10)      &  !990(3)    & !0.7(1)   & !996(3)   & -- & 0.020(3) & 75 & -- \\   
            & 4 & 930527 & 49134(22)      &  !!10(2)    & !0.6(2)   & !!13.2(13)& -- & --    & -- & -- \\ 
            & 5 & 940111 & 49363(130)     &  !142(21)   & !0.68(16) & !146(38)  & -- & --    & -- & -- \\
            & 6 & 940620 & 49523(17)$^*$  &  !!33(3)    &$-$0.55(9) & !!37(35)  & -- & --    & -- & -- \\
            & 7 & 950218 & 49766(2)$^*$   &  !!11(1)    &$-$0.26(6) & !!15(2)   & -- & --    & -- & -- \\
            & 8 & 950706 & 49904(16)      &  !!16(7)    &$-$1.9(4)  & !!31(1)   & -- & --    & -- & -- \\ 
            & 9 & 951018 & 50008(16)      & 1636(13)    & !3.3(4)   & 1648(3)   & -- & 0.004(1) & 300 & -- \\ 
            & 10 & 960826 & 50321.7(6)$^*$ &  !!27(1)    & !0.61(6)  & !!29.9(8)   & -- & --    & -- & -- \\
            & 11 & 970322 & 50528.9(8)$^*$ &  !!20(4)    & !1.0(4)   & !!23.4(5)   & -- & --    & -- & -- \\ 
            & 12 & 970823 & 50683(13)      &  !703(4)    & !1.2(3)   & !707.5(7)  & -- & --    & -- & -- \\ \\
1614$-$5047 & 1 & 950326 & 49803(16)      &  6456(56)   & !9.7(2)   & 6460(80) & --  & 0.538(11) & 2000    & 2168 \\ \\
1709$-$4428 & 1 & 920605 & 48778(34)      &  2012.3(2)  & !0.20(6)  & 2028(20) &--   & 0.133(7)  & 1420(90) & 16670  \\ \\
1731$-$4744 & 1 & 940204 & 49387.3(2)$^*$ & !135(1)     & !1.11(8)  & !139.2(6) & --  & 0.079(3)  & 263(23) & 1171   \\ 
            & 2 & 970912 & 50703(5)$^*$   & !!!2.6(6)   & !0.8(1)   & !!!3.1(5)   &--   & 0.25(13)  & 250    & --   \\ \\
1801$-$2304 & 1 & 910717 & 48454.1(3)$^*$ & !346(7)     & !1.5(9)   & !351(1)   & --       &--          &--      & 5061 \\
            & 2 & 941213 & 49701(1)$^*$   & !!64.7(5)   & !0.18(7)  & !!60.8(4)  & --       &--          &--     & --  \\
            & 3 & 951205 & 50050(5)$^*$   & !!22.6(6)   & !0.02(14) & !!17.0(5)  & --     &--        &--    & --   \\ 
            & 4 & 961125 & 50392(70)      & !!84(6)     & !1.7(8)   & !!87(2)    & -- &0.25(3)     & 100     & --   \\ \\
1801$-$2451 & 1 & 940504 & 49476(6)       & 1998(7)     & !4.85(28) & 1988(34) & --   &  0.188(12) & 800(60) & 5023  \\
            & 2 & 970722 & 50651(5)       & 1237(4)     & !3.87(9)  & 1248(9)  & --   &  0.202(6)  & 600      & --  \\ \\
1803$-$2137 & 1 & 971104 & 50765(15)      & 3200(27)    & 10.7(15 ) & 3185(25) & -- & 0.161(6) & 855(35) & 334 \\ 
            & 2 & 971104 & 50765(15)      &   --        &  --       & !!27(3)  & -- & 0.010(3) & 18(2)   & -- \\ \\  
\end{tabular}   
\label{tb:glparam}
$^*$ Glitch epoch determined by phase fit.
\end{minipage} 
\end{table*}

\subsection{PSR J0835$-$4510, PSR B0833$-$45, the Vela pulsar}
The Vela pulsar is well known to suffer many giant glitches
\cite{cdk88,mhmk90,lpsc96} and is being regularly monitored at a number of
observatories. At Parkes, we are unable to make such regular and frequent
observations. In this paper, we describe observations of the latest glitch
which has only been briefly reported \cite{fla96}.  

Table~\ref{tb:pparam} gives the results of fitting for $\nu$ and its first
two time derivatives, both before and after the glitch. The pre-glitch fit
was obtained from an approximately 1-year data span and shows a significant
$\ddot\nu$ resulting from previous glitches. The post-glitch fit given in
Table~\ref{tb:pparam} is for an 82-day span commencing 18 days after the
glitch. A timing model with polynomial terms up to $\ddot\nu$ is not a good
fit to longer data spans. 

\begin{figure} 
\centerline{\psfig{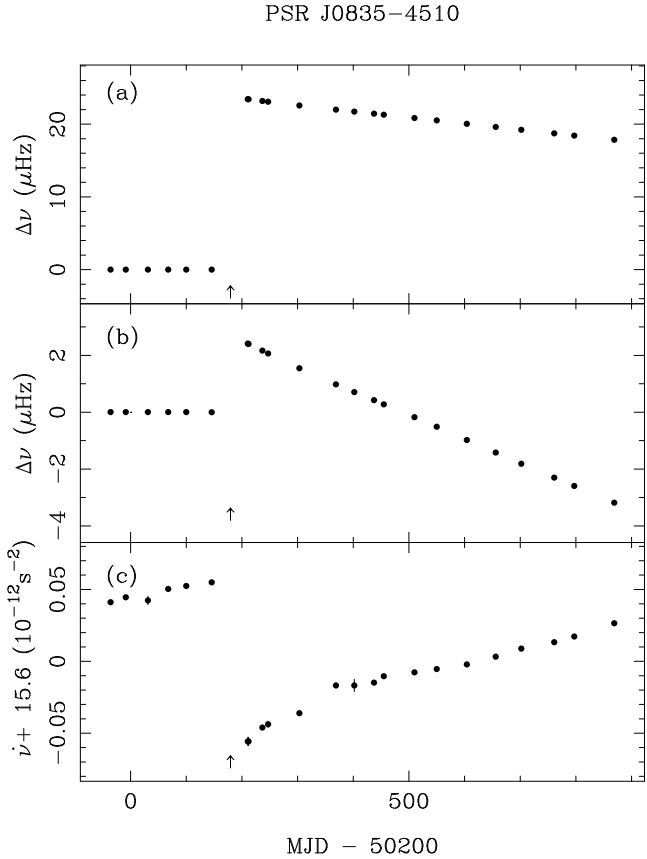}} 
\caption{The 1996 October glitch of PSR J0835$-$4510: variations of (a)
frequency residual $\Delta\nu$ relative to the pre-glitch solution, (b) an
expanded plot of $\Delta\nu$ where the mean post-glitch value has been
subtracted from the post-glitch data, and (c) the variations of
$\dot\nu$. The epoch of the glitch is indicated by a small arrow near the
base of each plot. }
\label{fg:0835nu}
\end{figure}

Fig.~\ref{fg:0835nu} shows the time dependence of the frequency residual
$\Delta\nu$ and of $\dot\nu$ around the time of the glitch. The $\Delta\nu$
values plotted are differences between the values of $\nu$ obtained from
independent fits to short sections of data, typically of span 20 -- 30 days,
and those determined from the predictions of the pre-glitch model given in
Table~\ref{tb:pparam}. This plot shows an approximately exponential recovery
in $\dot\nu$ after the glitch, followed by an approximately linear increase
in $\dot\nu$. This behaviour is similar to that seen in previous Vela
glitches \cite{lpsc96}.
 
Flanagan (1997) gives $\Delta\nu_g/\nu = 2.15(2) \times 10^{-6}$ and
an epoch for the glitch of 1996, October 13.394 UT, corresponding to
MJD 50369.394. Fractional changes in $\nu$ and $\dot\nu$ at the time
of the glitch, obtained by extrapolating the pre-glitch and
post-glitch fits (Table~\ref{tb:pparam}) to the glitch epoch are given
in the fifth and sixth columns of Table~\ref{tb:glparam}. Results of
fitting the exponential model (Equation~\ref{eq:glmodel}) to the
interval from 350 days before the glitch to 200 days after are also
given in Table~\ref{tb:glparam}. In this fit, $\nu$, $\dot\nu$ and
$\ddot\nu$ were held at their pre-glitch values
(Table~\ref{tb:pparam}). The small rms residual shows that the
exponential model with about 40 per cent of the initial glitch in
frequency decaying on a timescale of about 900 d is a very good
representation of the post-glitch relaxation, at least over the data
span fitted. Fitting of $\nu$, $\dot\nu$ and $\ddot\nu$ to the
post-glitch data instead of the three parameters of the exponential
fit gives a substantially worse fit with rms residual 360 $\mu$s and a
systematic quartic term in the residuals. For the exponential fit,
most of the rms residual comes from the first few post-glitch points
which are positive, indicating an additional short-term
recovery. However, it was not possible to fit for the parameters of
this. The fitted glitch epoch, given in the third column of
Table~\ref{tb:glparam}, is slightly earlier than that given by
Flanagan (1997); this is consistent with the presence of more rapid
post-glitch relaxation than modelled.

\subsection{PSR J1048$-$5832, PSR B1046$-$58}
\begin{figure} 
\centerline{\psfig{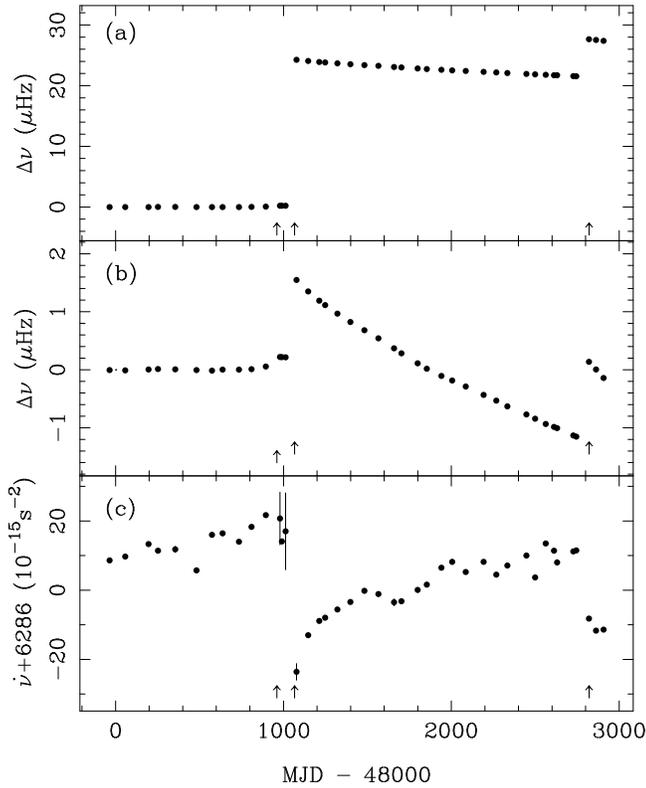}} 
\caption{Glitches of PSR J1048$-$5832: variations of (a) frequency residual
$\Delta\nu$ relative to the pre-glitch solution, (b) an expanded plot of
$\Delta\nu$ where, for the second and third glitches, the mean residual up
to the next glitch, or to the end of the data if there is no following
glitch, has been removed from the corresponding interval (indicated by the
raised arrows), and (c) the variations of $\dot\nu$. }
\label{fg:1048nu}
\end{figure}

\begin{figure} 
\centerline{\psfig{file=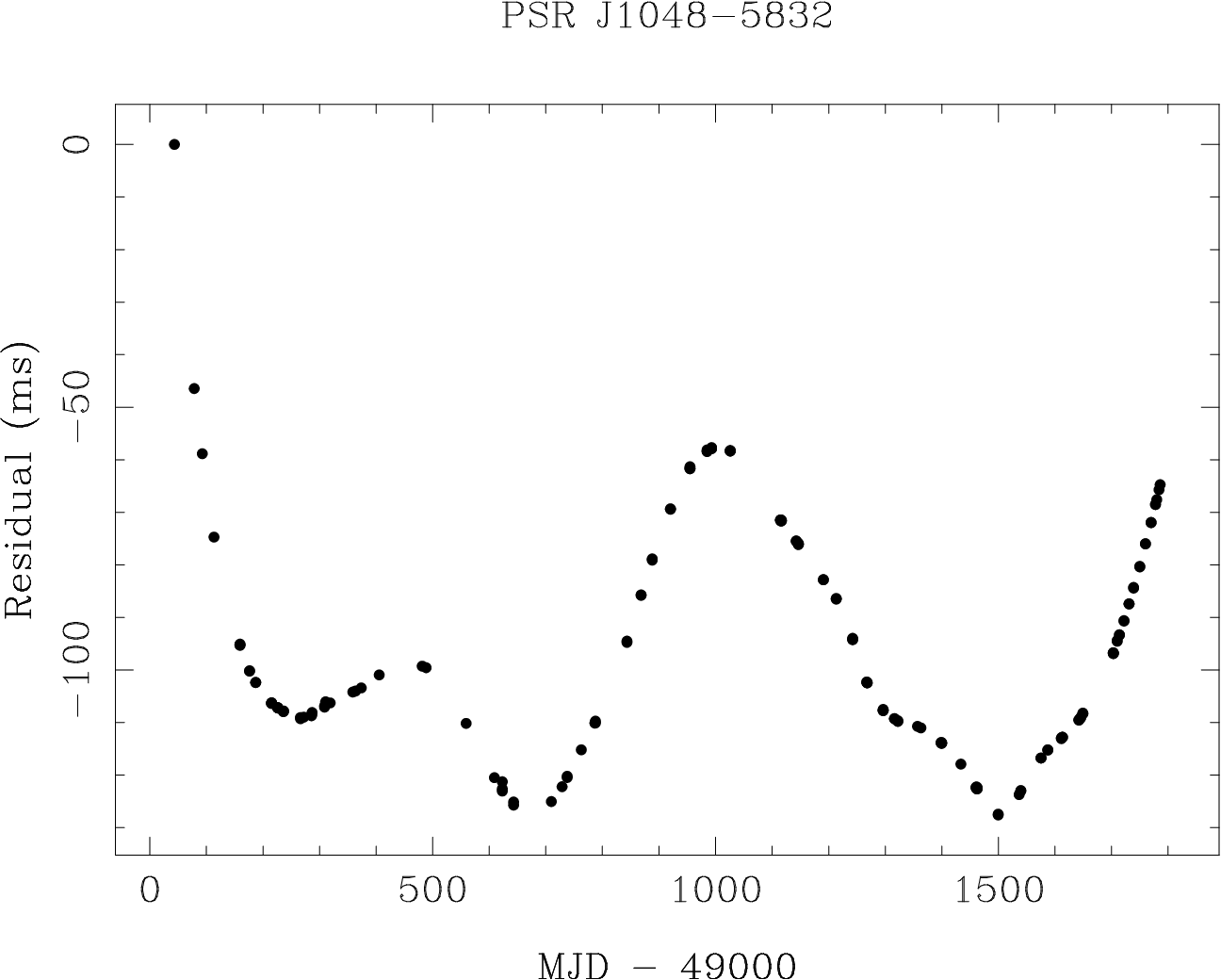,width=85mm,angle=270}} 
\caption{Timing residuals for PSR J1048$-$5832 between the two large glitches.}
\label{fg:1048res2}
\end{figure}

This young pulsar was discovered in a high-frequency survey of the southern
Galactic plane \cite{jlm+92}. Evidence for gamma-ray pulsations has been
found by Kaspi et al. (2000).\nocite{klm+00} The pulsar suffered two large
glitches within the observed data span, the first of size $\Delta\nu_g/\nu
\sim3\times10^{-6}$ in 1993 February (MJD $\sim 49034$) and the second of
size $\sim 0.7\times10^{-6}$ in 1997 December (MJD $\sim 50788$). Observed
frequency residuals are shown in Fig.~\ref{fg:1048nu}. The expanded plot of
$\Delta\nu$ (Panel b) shows a third and much smaller glitch about
100~d before the large first glitch. Fits to the inter-glitch intervals
(Table~\ref{tb:pparam}) give evidence for large fluctuations in the
period. Fig.~\ref{fg:1048res2} gives timing residuals for the data span
following the largest glitch, showing quasi-random fluctuations with a
timescale of a few hundred days.

Table~\ref{tb:glpsrs} gives an improved position for PSR J1048$-$5832 derived
from the data following the largest glitch. Most of the systematic oscillation
was removed by fitting up to the twelfth pulse frequency derivative at the
same time as the position was determined. A fit to the whole data set, including
parameters for the glitches, gave the same position within the combined
errors. This position differs by 8\arcsec~ from that given by Johnston et
al. (1995); this difference can be attributed to the different sets and the
presence of the strong period irregularities. Recently, Stappers et
al. (1999)\nocite{sgj99} have used the Australia Telescope Compact Array to
determine an interferometric position for the pulsar: R.A. (J2000) 10$^{\rm h}$
48$^{\rm m}$ $12\fs604 \pm 0\fs008$, Dec. (J2000) $-58\degr~32\arcmin~03\farcs75 \pm
0\farcs05$. This position agrees with the timing position in R.A., but
differs by $2\farcs45 \pm 0\farcs80$ in declination, probably as a result of
unmodelled period irregularities affecting the timing position.

Extrapolation of fits to the data sets on either side of the
glitches gives the estimates of glitch parameters listed in columns 5 and 6 of
Table~\ref{tb:glparam}. Glitch parameters from TOA fits
are given in the remaining columns.  Because of the contaminating effect of
the period fluctuations, the parameters of the large glitches were determined
by fitting the data span 100 -- 150 days before and after each glitch. The
effects of the earlier small glitch were determined separately by fitting
the interval between 80 days prior to it and up to the large glitch, and
were subtracted from the fit to the large glitch. Parameters from these fits
are given in Table~\ref{tb:glparam}. Because of the large systematic period
variations, it is not possible to reliably measure the
post-glitch decay times. Fig.~\ref{fg:1048res2} gives some evidence for a
relaxation with a timescale of $\sim 100$~d following the 1993 glitch and
somewhat longer for the 1997 glitch. Setting the decay time to 100~d and
400~d respectively for these two glitches gave the parameters listed in
Table~\ref{tb:glparam}. The derived $Q$ for the 1997 glitch appears
significantly larger than that for the 1993 glitch, but this result is
uncertain given the uncertainty in the relaxation time and the short data
span following the glitch.

\subsection{PSR J1105$-$6107}
PSR J1105$-$6107 is a young pulsar with the relatively short period of 63 ms
discovered in 1993 using the Parkes radio telescope \cite{kbm+97}. It is
located close to but outside the boundaries of the supernova remnant
G290.1$-$0.8 and within the error box for the EGRET gamma-ray source 2EG
J1103$-$6106 \cite{klm+00}. Its association with these objects remains
plausible but unproven. Kaspi et al. (1997) reported on timing observations
from 1993 July to 1996 July. Here we extend this interval to the end of 1998,
and show that there were two glitches, a large one around the end of
November, 1996 (MJD $\sim 50417$) , and a small one about 200 days later.

\begin{figure} 
\centerline{\psfig{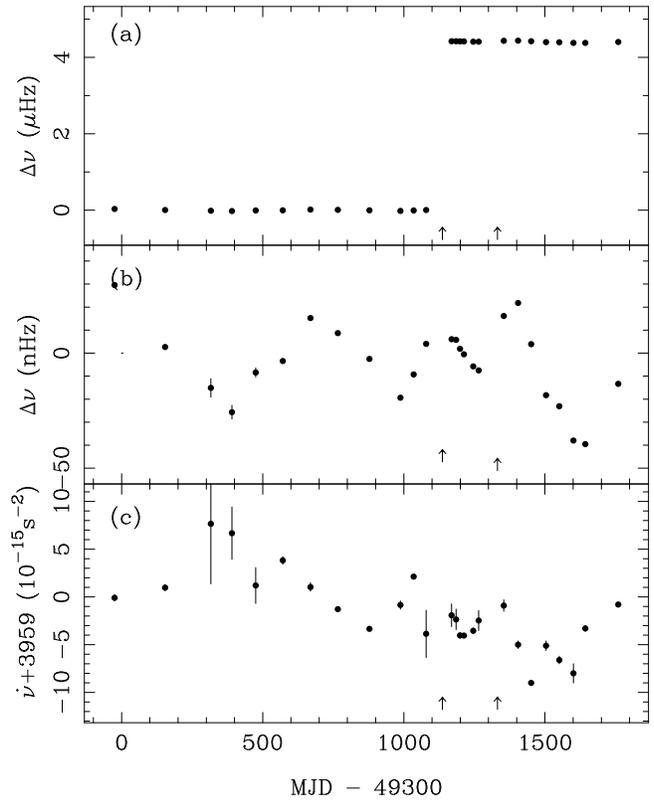}} 
\caption{Glitches of PSR J1105$-$6107: variations of (a) frequency residual
$\Delta\nu$ relative to the pre-glitch solution, (b) an expanded plot of
$\Delta\nu$ where the mean residual between the large glitch and the second
glitch has been removed from the data following the large glitch (indicated
by the raised arrow), and (c), the variations of $\dot\nu$. }
\label{fg:1105nu}
\end{figure}

\begin{figure} 
\centerline{\psfig{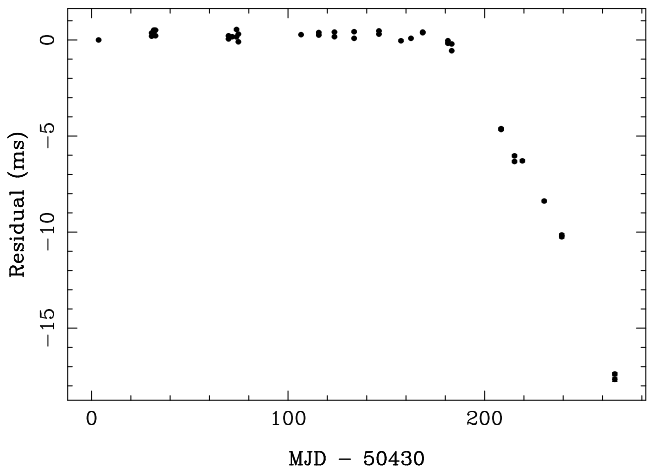}} 
\caption{Timing residuals around the time of the small glitch in the
period of PSR J1105$-$6107. }
\label{fg:1105_gl2}
\end{figure}

Fig.~\ref{fg:1105nu} gives the variations in pulse frequency and frequency
derivative over the observed data span, clearly showing the larger glitch
which is of fractional size $\Delta\nu_g/\nu \sim 0.28 \times
10^{-6}$. There was a gap between bracketing observations of about 30 days,
so the glitch epoch is not well determined. The expanded plot (b) shows
that, despite the substantial period irregularities, there is good evidence
for a small glitch about 200 days after the large
glitch. Fig.~\ref{fg:1105_gl2} shows phase residuals around the time of this
small glitch. Although this glitch is by far the smallest discussed in this
paper, there is little doubt about its reality. Another small glitch may
have occurred near the end of the data set, but there are insufficient
post-glitch observations to confirm this. The position given in
Table~\ref{tb:glpsrs} was determined from a fit to the data prior to the
large glitch, with seven frequency derivatives to absorb the period
irregularities.

Table~\ref{tb:pparam} gives fits to the three data spans delimited by the
data set and the two glitches; the position was held at the value quoted in
Table~\ref{tb:glpsrs}. The second derivative terms in the first and third
fits are dominated by random period irregularities.  Glitch parameters
obtained by extrapolating these fits and from phase fits across the glitches
are given in Table~\ref{tb:glparam}. Because of the period
noise, the phase fits were restricted to intervals of 150 days before the
first glitch, the inter-glitch interval, and 150 days after the second
glitch. The fractional increase $\dot\nu$ after the first glitch is small,
implying a similarly small value of the ratio $Q/\tau_d$ (Equation
\ref{eq:glnudot}). Because of the small data span, $\tau_d$ is not well
determined, but taking 100 d gives $Q \sim 0.035$; i.e., only a few percent
of the glitch is likely to decay. This is confirmed by a fit to the whole
data span including both glitches. 

\subsection{PSR J1123$-$6259}
PSR J1123$-$6259, discovered in the Parkes southern pulsar survey
\cite{mld+96}, has a period of 271 ms and a modest period derivative, giving
a characteristic age of $\sim 8 \times 10^5$ yr. Despite this relatively large
characteristic age, a glitch was detected between 1994 December 17 -- 22
(MJD $\sim 49706$). The pulsar position was determined from the post-glitch
data (Table~\ref{tb:glpsrs}); no pre-whitening was necessary as random
period irregularities are small in this pulsar. Fits to the pre- and
post-glitch data with this position are given in
Table~\ref{tb:pparam}. Extrapolation of these fits to a central epoch
(Table~\ref{tb:glparam}) shows that the glitch was of magnitude $\Delta\nu /
\nu \sim 7.5 \times 10^{-7}$. \footnote{A glitch of this magnitude was
reported by D'Amico et al. (1998)\nocite{dsb+98} to have occurred at MJD
$48650 \pm 20$ (1992 December). This epoch is in error. It was in fact
the December 1994 glitch discussed in this paper.}

\begin{figure} 
\centerline{\psfig{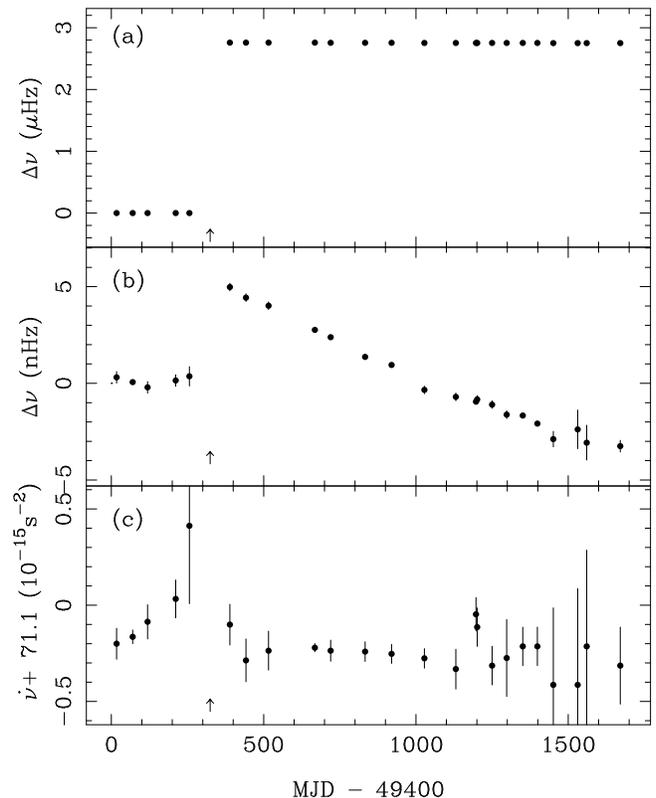}} 
\caption{The 1994 December glitch of PSR J1123$-$6259: variations of (a)
frequency residual $\Delta\nu$ relative to the pre-glitch solution, (b) an
expanded plot of $\Delta\nu$ where the mean post-glitch value has been
subtracted, and (c) the variations of $\dot\nu$.}
\label{fg:1123nu}
\end{figure}

As Table~\ref{tb:glparam} and Fig.~\ref{fg:1123nu} show, there was a small
but significant increase in $|\dot \nu|$ at the time of the glitch. A fit to
the entire data span including glitch parameters but with the position and
pre-glitch frequency parameters held to the values given in
Table~\ref{tb:glpsrs} and Table~\ref{tb:pparam} is an excellent fit to the
data and gives an unambiguous value for the epoch of the glitch with an
uncertainty of about 15 min on the assumption that there was phase
continuity across the glitch. The derived value of $Q$ is very small and
the decay time is long (Table~\ref{tb:glparam}).

\subsection{PSR J1341$-$6220}
PSR J1341$-$6220 is a young pulsar (characteristic age $\sim 12,000$ y)
discovered by Manchester et al. (1985)\nocite{mdt85} and associated with the
supernova remnant G308$-$0.1 by Kaspi et al. (1992).\nocite{kmj+92} Kaspi et
al. reported two large glitches ($\Delta\nu_g /\nu \sim 1.5 \times 10^{-6}$
and $\sim 1.0 \times 10^{-6}$) and one smaller one ($\sim 2.3 \times
10^{-8}$) within the period 1990 January to 1992 May, making this one of the
most actively glitching pulsars known. In this paper, we reanalyse the data
from 1990 to 1992 and extend the data set 1998 March. Unfortunately,
observations after this time were too sparse to permit unambiguous pulse
counting.

The position given in Table~\ref{tb:glpsrs} was determined in a simultaneous
fit across the whole data set, including all glitches (see below). It has
smaller estimated uncertainties than the position given by Kaspi et al. (1992), and
lies 4\farcs3 north of it. Given the prominence of period irregularities in
this object, this difference, while twice the combined uncertainties, is of
marginal significance. Determination of the DM is
complicated by the strong scattering tail shown by this pulsar, even at
frequencies around 1400 MHz. The value given in Table~\ref{tb:glpsrs} was
determined using TEMPO with data recorded between 1998 March and December at
frequencies close to 1400 MHz and 1700 MHz. The relative phase of the
standard profiles at the two frequencies was adjusted to allow for the
effects of scattering. The derived DM is much more accurate than the value
given by Kaspi et al. (1992) and just outside the error range of this value.

\begin{figure} 
\centerline{\psfig{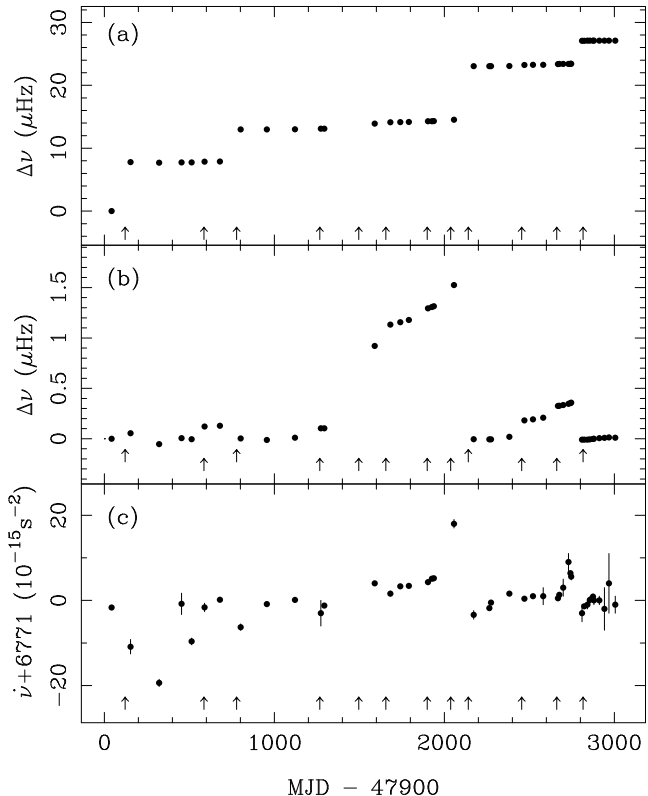}} 
\caption{The glitches of PSR J1341$-$6220: variations of (a) frequency
residual $\Delta\nu$ relative to the pre-glitch solution, (b) an expanded
plot of $\Delta\nu$ where the mean residual between glitches with a raised
arrow and the following glitch has been removed from data after the marked
glitch, and (c) the variations of $\dot\nu$.}
\label{fg:1341nu}
\end{figure}

Fig.~\ref{fg:1341nu} shows the very complicated period history of this
pulsar. A total of 12 glitches were observed in the 8.2-year interval,
making this the most frequently glitching pulsar known, with a mean glitch
interval of 250 days. The next most frequent is PSR B1737$-$30, with a mean
interval of 345 days \cite{sl96}. Four of the glitches were relatively
large, with $\Delta\nu_g /\nu > 0.7 \times 10^{-6}$ and for two the fractional
glitch size exceeds $10^{-6}$. Most of the other glitches are quite small,
with all except one having $\Delta\nu_g /\nu < 0.04 \times 10^{-6}$. They are,
however, unambiguous on phase-time plots.

For most of the glitches, there is little evidence for any post-glitch
relaxation. This is partly because of the relatively poorly sampled data,
especially in the earlier years, but also because of the short time between
successive glitches. For the third and ninth glitches, there is weak evidence
in the $\dot\nu$ plot for some relaxation. 

Table~\ref{tb:pparam} gives fits to the inter-glitch intervals. Because of
the limited data span prior to the first observed glitch, the value of
$\dot\nu$ was held fixed at a value close to the long-term mean for this
fit. For about half of the intervals, there was a clear $\ddot\nu$ term in
the residuals and this term was fitted. 

Changes in $\nu$ and $\dot\nu$ at the time of each glitch, obtained by
extrapolation of these fits to the glitch epoch, are given in
Table~\ref{tb:glparam}. This table also gives glitch parameters from a
single fit to the whole data set, solving simultaneously for the pulsar
position, pulsar frequency ($\nu$), mean frequency derivative
($\dot\nu$), and the parameters for all 12 glitches. As mentioned
above, for most of the glitches, there was no significant post-glitch
relaxation. For the third and ninth glitches, a post-glitch exponential
decay was fitted; in both cases the decay time constant was not fitted for,
but was determined by trial to minimise the final residual. In both cases
the derived $Q$ values are relatively small.

\subsection{PSR J1614$-$5047, PSR B1610$-$50}
\begin{figure} 
\centerline{\psfig{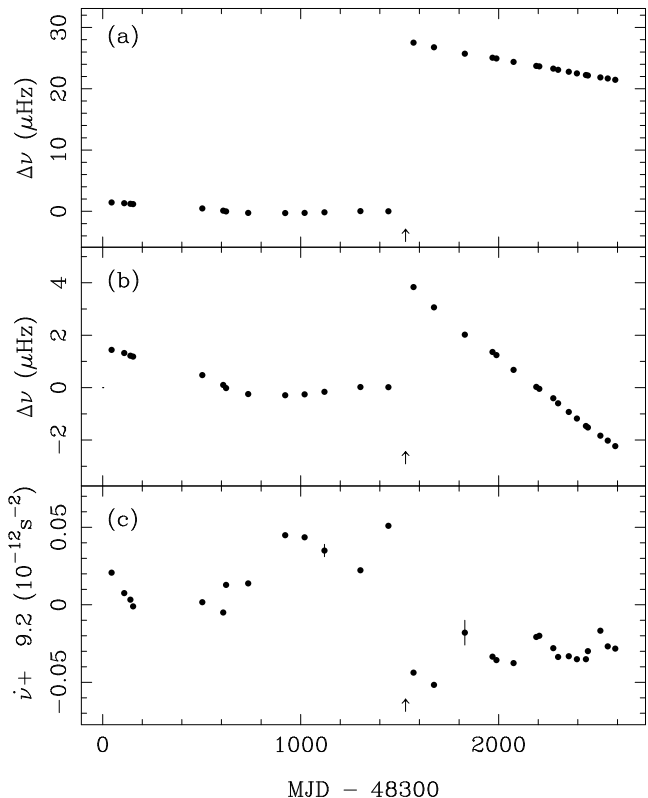}} 
\caption{The 1995 June glitch of PSR J1614$-$5047: variations of (a)
frequency residual $\Delta\nu$ relative to the pre-glitch solution, (b) an
expanded plot of $\Delta\nu$ where the mean residual after the glitch has
been removed, and (c) the variations of $\dot\nu$.}
\label{fg:1614nu}
\end{figure}

This pulsar has a period of 231 ms and a very large period derivative ($495
\times 10^{-15}$), implying a small characteristic age of $\sim 7400$
yr. In terms of its period irregularities, it is one of the noisiest pulsars
known; this makes it difficult at times to keep track of the pulse
phase. Since its discovery in late 1989 \cite{jlm+92}, there has been no
clear evidence for a glitch although, particularly where there is a
significant gap in the timing data, it is often difficult to distinguish
between a (small) glitch and more continuous period irregularities. However,
in 1995 June (MJD $\sim 49802$), there was a massive glitch with
$\Delta\nu_g/\nu \sim 6.5 \times 10^{-6}$, the largest ever observed in any
pulsar (cf. Shemar \& Lyne 1996).  Fig.~\ref{fg:1614nu} shows this glitch
and also illustrates the more continuous period irregularities.

These irregularities make it difficult to determine the pulsar position from
timing data. Previously published positions have differed by much more than
the quoted uncertainties \cite{tml93,jml+95}. Taking the data span from MJD 50269
to to 50778 (which is free of major irregularities -- see below) and fitting
for position and two frequency derivatives gives the position quoted in
Table~\ref{tb:glpsrs}. An independent fit to the MJD range 48732 -- 49093
gave a position with uncertainty of about 2\arcsec, consistent with the
position from the longer data span. Subsequent fits were made keeping the
position fixed at the Table~\ref{tb:glpsrs} value. Stappers et al. (1999)
have recently determined an interferometric position for this pulsar:
R.A. (J2000) 16$^{\rm h}$ 14$^{\rm m}$ $11\fs55 \pm 0\fs01$, Dec. (J2000)
$-50\degr~48\arcmin~01\farcs9 \pm 0\farcs1$. As for PSR J1048$-$5832, the
differences most probably result from period irregularities affecting the
timing position.

The slope of the post-glitch data in the top two plots indicates a large
change in frequency derivative at the time of the glitch. Because of the
strong irregularities in the period, the pre- and post-glitch fits given in
Table~\ref{tb:pparam} are restricted to intervals of less than 300
days. Extrapolation of those fits gives the glitch parameters in columns 5
and 6 of Table~\ref{tb:glparam}. Glitch parameters were also obtained from a
single timing solution over the same interval as the pre- and post-glitch
fits (MJD 49559 -- 50117) and are given in columns 7 -- 10 of
Table~\ref{tb:glparam}. This fit did not converge well when fitting for the
glitch decay time, so this was held fixed at 2000 d, a value representative
of those obtained from the fitting. From the post-fit residuals, it is clear
that there is also a more rapid decay of a portion of the glitch with a
timescale of 10 -- 20 days. There is good agreement between the two methods
of deriving the glitch parameters. From Equation~\ref{eq:glnudot}, the $Q$
and $\tau_d$ values given in Table~\ref{tb:glparam} imply a fractional change
in $\dot\nu$ of 0.0096, compared to 0.0097 from the
extrapolation. Unfortunately, because of the large gap between the
observations bracketing the glitch and the large size of the glitch, it is
not possible to accurately determine the glitch epoch from phase continuity. 

This is the largest glitch yet observed. The previous largest was
for PSR B0355+54 for which $\Delta\nu/\nu$ was $\sim 4.37 \times 10^{-6}$
\cite{lyn87}. There was very little decay of the frequency step in PSR
B0355+54, whereas the $Q$ value for the PSR J1614$-$5047 glitch indicates
that more than half of it will decay away over a few years. In this respect,
the PSR J1614$-$5047 glitch is more similar to the giant glitches in the Vela
pulsar, which decay by a similar amount over similiar timescales.

\begin{figure*}
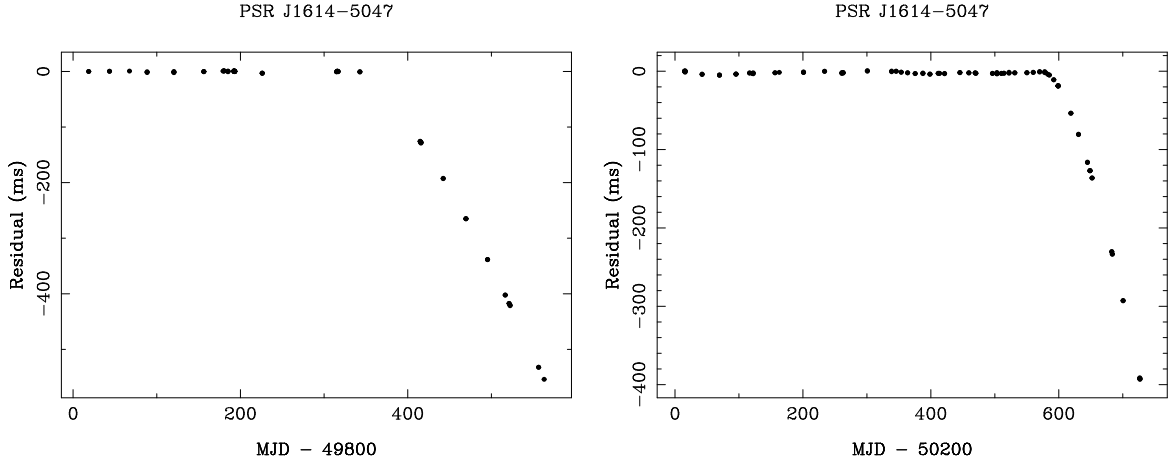
 
\begin{tabular}{ll} 
\mbox{\psfig{file=1614_gl2_res.ps,width=75mm,angle=270}} &
\mbox{\psfig{file=1614_gl3_res.ps,width=75mm,angle=270}}
\end{tabular}
\caption{Variations of timing phase residuals around two epochs of rapid
frequency change in PSR J1614$-$5047}
\label{fg:1614_phs}
\end{figure*}

Phase fitting of post-glitch data revealed two further abrupt changes in
pulse frequency. Phase plots for the intervals around these events are shown
in Fig.~\ref{fg:1614_phs}. These phase plots have the character of glitch
events, with a persistent fractional frequency change $\Delta\nu_g / \nu \sim
0.03 \times 10^{-6}$ at about MJD 50170 and 50780, respectively. The first
event may indeed be a small glitch -- it is not possible to tell because of
the large data gap -- but the second is not. This frequency change is
resolved, taking place over about 10 days. In both cases, there appears to
be a significant decrease in $|\dot\nu|$ associated with the event.  These
rapid frequency changes could be classed as just part of the frequency
irregularities which are prominent in this pulsar, but in general these
irregularities have a much longer timescale and are smaller in
amplitude. This is demonstrated by the small residuals for the fits to data
before these events; for the second event, the fit includes only two frequency
derivatives and has an rms residual of only 1.2 ms
(Table~\ref{tb:pparam}). The MJD 50780 event stands out from the general
irregularities with a much larger rate of frequency change over the 10 days.
Phase fits to the post-glitch data given in Table~\ref{tb:pparam} have been
split into three sections to avoid contamination by these sudden frequency
changes. The last ends at MJD 50926 (1998 April) since data are sparse after
that point and it is not possible to fit across the gaps without
ambiguity. The frequency second derivative terms are very significant in
these fits, but differ greatly in both sign and magnitude for the different
segments. They are clearly related to the period-noise processes occurring
in this star and not to the secular slowdown.

\subsection{PSR J1709$-$4428, PSR B1706$-$44}
This pulsar, discovered by Johnston et al. (1992),\nocite{jlm+92} is of
interest for several reasons. It is young ($\tau_c \sim 17$ kyr) and has the
third highest known value of $\dot E / d^2$, where $\dot E$ is the spin-down
luminosity and $d$ is the pulsar distance, after the Crab and Vela
pulsars. It has been detected at gamma-ray wavelengths \cite{tab+92,tbb+96},
X-ray wavelengths \cite{bbt95} and possibly at TeV energies
\cite{kto+95}. A supernova remnant association was proposed by McAdam,
Osborne \& Parkinson (1993),\nocite{mop93} but deemed unlikely by Nicastro,
Johnston \& Koribalski (1996)\nocite{njk96}. 

Johnston et al. (1995) presented results from two years of Parkes timing
observations showing that the pulsar suffered a giant glitch of magnitude
$\Delta\nu_g / \nu \sim 2 \times 10^{-6}$ near the end of May, 1992 ($\sim$
MJD 48778). In this paper we analyse data from 1990 January to 1998
December. To determine the pulsar position, we have taken data from 150 days
after the glitch (MJD 48928 -- 51155) and fitted nine frequency derivatives
to absorb the period irregularities. This fit had an rms residual of only
211 $\mu$s and gave the position listed in Table~\ref{tb:glpsrs}. The
derived position is about $12''$ southeast of the position given by Johnston
et al. (1995), but agrees better with an interferometric position given by
Frail \& Scharringhausen (1997)\nocite{fs97}: R.A. (J2000) 17$^{\rm h}$
09$^{\rm m}$ $42\fs75 \pm 0\fs01$, Dec. (J2000)
$-44\degr~29\arcmin~06\farcs6 \pm 0\farcs3$. The timing fits described below
were obtained with the position held at the Table~\ref{tb:glpsrs} value.

\begin{figure} 
\centerline{\psfig{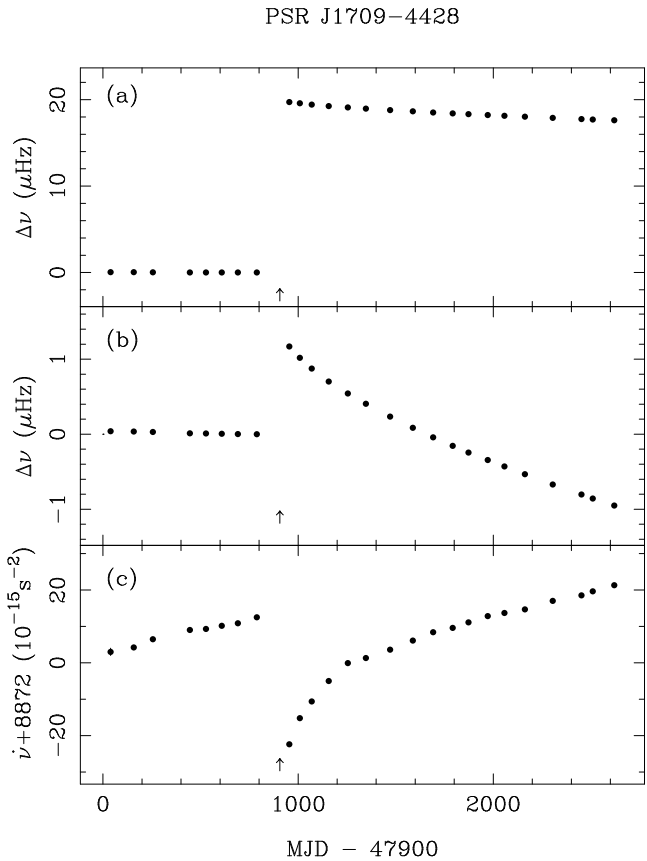}} 
\caption{The 1992 May glitch (MJD 48778) of PSR J1709$-$4428: variations of (a)
frequency residual $\Delta\nu$ relative to the pre-glitch solution, (b) an
expanded plot of $\Delta\nu$ where the mean residual after the 
glitch has been removed, and (c) the
variations of $\dot\nu$. }
\label{fg:1709nu}
\end{figure}

\begin{figure} 
\centerline{\psfig{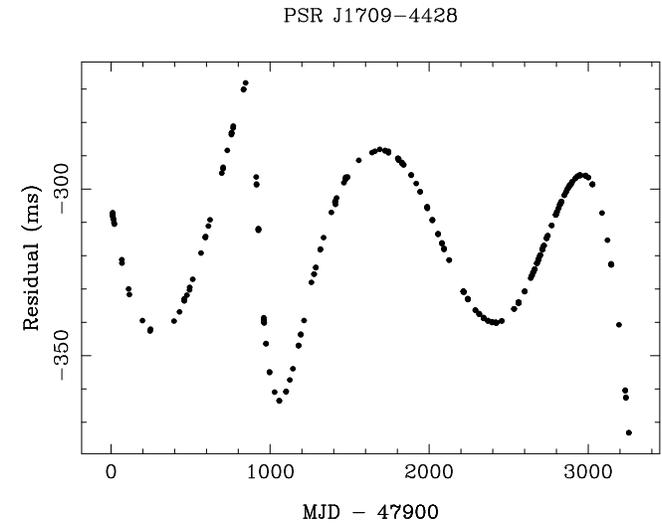}} 
\caption{Post-fit timing residuals for a fit to the entire data set for PSR
J1709$-$4428, including the glitch at MJD 48778}
\label{fg:1709_g1_res}
\end{figure}

Fig.~\ref{fg:1709nu} shows the large jump with a classic exponential
recovery on a timescale of $\sim 100$ days coupled with a longer term
relaxation. Fits of a cubic phase polynomial to the pre-glitch data, to the
150 days after the glitch and to data from 150 days post-glitch are given in
Table~\ref{tb:pparam}. The post-glitch fit has large systematic residuals,
dominated by the post-glitch recovery. Glitch parameters obtained by
extrapolating the first and second fits to the glitch epoch are given in
Table~\ref{tb:glparam} along with those from a phase fit to the whole data
set. This latter fit is strongly affected by the strong period
irregularities (Fig.~\ref{fg:1709_g1_res}) and so the derived parameters are
only approximate. There is clear evidence in the post-fit residuals for a
more rapid exponential decay with time constant of the order of 100
days. However, it was not possible to fit for the parameters of this owing
to the contaminating effect of the period irregularities.

\subsection{PSR J1731$-$4744, PSR B1727$-$47}
\begin{figure} 
\centerline{\psfig{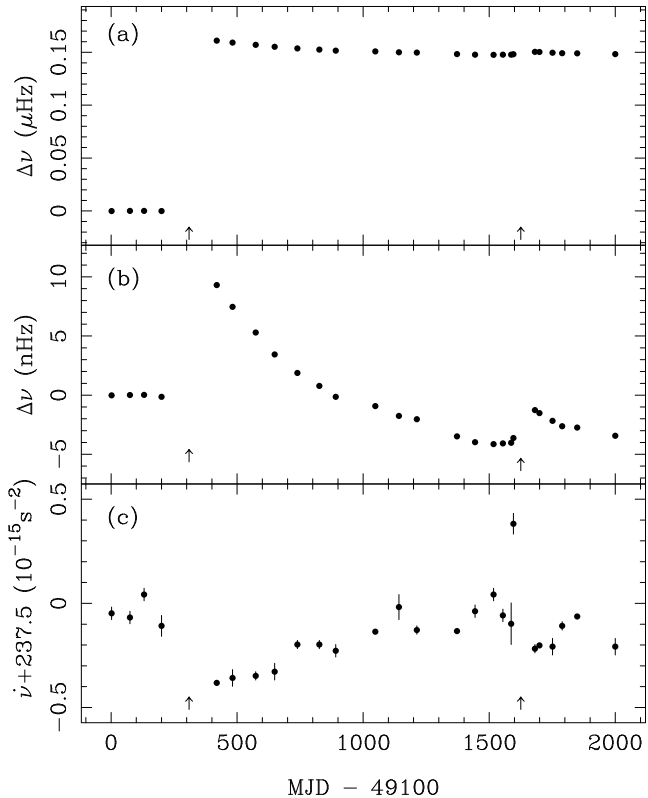}} 
\caption{Glitches of PSR J1731$-$4744: variations of (a) frequency residual
$\Delta\nu$ relative to the pre-glitch solution, (b) an expanded plot of
$\Delta\nu$ where the mean residual in the inter-glitch interval has been
removed from the following data, and (c) the variations of $\dot\nu$. }
\label{fg:1731nu}
\end{figure}

PSR J1731$-$4744 is a long-period pulsar (0.830 s) which
has a large period derivative giving a relatively low characteristic age,
$\sim 80,000$ years. D'Alessandro \& McCulloch (1997) observed this pulsar
at Mt Pleasant Observatory in Tasmania from 1987 to 1994 and detected a
relatively large glitch at MJD 49387.68 (1994 February 2) but found no
evidence for any recovery in 10 months of observations after the glitch. The
Parkes observations extend to the end of 1998 (Fig.~\ref{fg:1731nu}) and
show a clear recovery from this glitch which is roughly exponential. They
also reveal a second smaller glitch in 1997 September which has a similar
recovery.

Table~\ref{tb:glpsrs} gives an improved position and DM for this pulsar. The
position was determined from data between the two glitches by fitting for
the position and five frequency-derivative terms to absorb the period
irregularities. The DM was determined by fitting to data in the MJD range
49400 to 49900 where there were observations around 430 MHz as well as
around 1400 MHz. This value is significantly different from the best
previously published value ($121.9 \pm 0.1$ cm$^{-3}$ pc; McCulloch et
al. 1973),\nocite{mka+73} most probably reflecting a changing line-of-sight
path through the interstellar medium. The average rate of DM change over the
24 years is $\sim 0.6$ cm$^{-3}$ pc y$^{-1}$. DM changes have been observed
in other pulsars, notably the Crab \cite{ir77} and Vela \cite{hhc85}
pulsars. The average rate of change of DM for
PSR J1731$-$4744 is comparable to that observed for the Vela pulsar and
larger than that for the Crab pulsar. The large changes observed for the
these two pulsars are attributed to ionised gas within the associated
supernova remnant, but PSR J1731$-$4744 has no associated supernova remnant.

Fits to the inter-glitch intervals are given in Table~\ref{tb:pparam}. These
fits represent the data reasonably well, except that there is a significant
quartic term in the residuals for the middle interval. This is mostly due to the
period irregularities which are evident in Fig.~\ref{fg:1731nu}.

Table~\ref{tb:glparam} gives parameters for the two glitches. The fitted
parameters were determined by a simultaneous fit to both
glitches. As mentioned above, a recovery after both glitches is clearly seen
in Fig.~\ref{fg:1731nu}, and these are well fitted by the exponential
model. For the first glitch, about 8 per cent of the glitch is recovered
with a time constant of about 260 days. This implies a fractional increment in
$\dot\nu$ at the time of the glitch of $2.25 \times 10^{-3}$. The glitch
parameters are in reasonable agreement with those quoted by
D'Alessandro \& McCulloch (1997); the differences of a few times the combined
uncertainties could result from the different models used for the extrapolation to
the time of the glitch. The data span following the second
glitch is too short to permit solving for the time constant, so 250 days was
assumed, giving a $Q$ value of about 25 per cent.

\subsection{PSR J1801$-$2304, PSR B1758$-$23}
This pulsar was detected in a search for short-period
pulsars associated with supernova remnants \cite{mdt85}. Despite its
relatively long period of 415 ms, this pulsar has a
short characteristic age (58 kyr). It lies close to the supernova remnant
W28, but its association with the remnant remains controversial
\cite{klm+93,fkv93}. It has the highest known dispersion measure (1074
cm$^{-3}$ pc) and even at 1.4 GHz the profile is very scattered, reducing the
precision of pulse timing observations at this and lower
frequencies. Furthermore, the pulsar lies very close to the ecliptic plane
and suffers frequent glitches, so that the pulsar position is not well determined
by pulse timing; the position quoted in Table~\ref{tb:glpsrs} and used for
the timing analyses is from the VLA observations of Frail et al. (1993).

\begin{figure} 
\centerline{\psfig{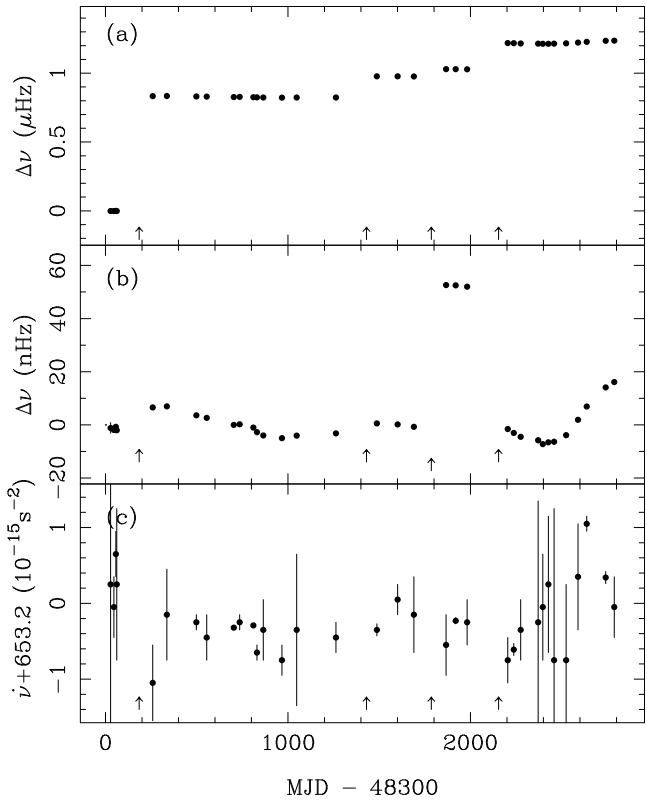}} 
\caption{Glitches of PSR J1801$-$2304: variations of (a) frequency residual
$\Delta\nu$ relative to the pre-glitch solution, (b) an expanded plot of
$\Delta\nu$ where the mean residual following each glitch with a raised arrow has been
removed, and (c) the variations of $\dot\nu$. }
\label{fg:1801-2304nu}
\end{figure}

Four glitches in this pulsar occurring between 1986 and the end of 1994 were
reported by Kaspi et al. (1993) and Shemar \& Lyne (1996). In this paper, we
present data on the last two of these glitches and two further glitches,
both of relatively small amplitude, occurring in 1995 and 1996
respectively. Fig.~\ref{fg:1801-2304nu} shows the variations in pulse
frequency and frequency derivative around these four glitches relative to
the pre-glitch solution given Shemar \& Lyne (1996) for the MJD 48454
glitch. This pre-glitch solution was adopted because that given in
Table~\ref{tb:pparam} is based on a data span of only 120 days and has a
significantly different value of $\dot\nu$ from the other fits, probably as
a result of intrinsic period irregularities. Fits to the inter-glitch
intervals given in Table~\ref{tb:pparam} have relatively large residuals
because of these irregularities. Only the fit to the data following the
fourth glitch required a $\ddot\nu$ term, indicating a significant
relaxation following the glitch.

Parameters for the four glitches are given in
Table~\ref{tb:glparam}, both from extrapolation of the polynomial fits given
in Table~\ref{tb:pparam} and from a simultaneous fit to the TOA data across
all four glitches. Values of $\Delta\nu/\nu$ and $\Delta\dot\nu/\dot\nu$ for
the first two jumps have smaller uncertainties but are consistent with the values
given by Shemar \& Lyne (1992). For the first three jumps the glitch epoch
is determined from the requirement of phase continuity over the glitch; for
the fourth glitch the data gap is too large for an unambiguous determination
of the glitch epoch. Consistent with the observation of a significant
$\ddot\nu$ term, post-fit residuals were significantly reduced by allowing
an exponential relaxation following the fourth jump. However, the presence
of period irregularities made fitting for the decay time impossible; the
value of 100 days was found by trial to give the best representation of the
data. 

\subsection{PSR J1801$-$2451, PSR B1757$-$24}
PSR J1801$-$2451 is a young pulsar ($\tau_c \sim 15,000$ y)
which is located at the ``beak of the Duck'', in the small nebula associated
with the much larger supernova remnant G5.4$-$1.2 \cite{mkj+91,fk91}. A
giant glitch in the pulse period was observed around MJD 49476 by Lyne et
al. (1996).\nocite{lkb+96} About a year of post-glitch data to MJD 49950
(1995 August) suggested a relaxation of a small fraction of the glitch
($Q \sim 0.005$) with a characteristic timescale of about
40 days.

\begin{figure} 
\centerline{\psfig{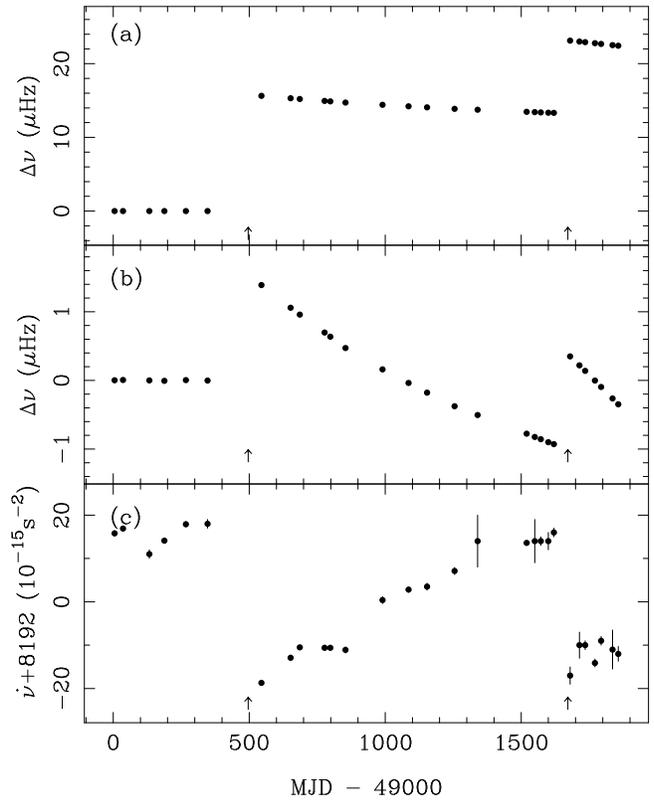}} 
\caption{Glitches of PSR J1801$-$2451: variations of (a) frequency residual
$\Delta\nu$ relative to the pre-glitch solution, (b) an expanded plot of
$\Delta\nu$ where the mean residual after each glitch has been
removed, and (c) the variations of $\dot\nu$. }
\label{fg:1801-2451nu}
\end{figure}

In Fig.~\ref{fg:1801-2451nu} we show observed pulse frequency variations
based on Parkes data from 1992 October to 1998 December. These show the
glitch observed by Lyne et al. (1996) and also a second smaller, but still
large, glitch around 1997 July. Polynomial fits to the inter-glitch
intervals given in Table~\ref{tb:pparam} have large residuals owing to the
presence of period irregularities. Extrapolation of the first two of these
fits to the glitch epoch determined by Lyne et al. (1996) gives results
consistent with those obtained by these authors, with $\Delta\nu_g/\nu \sim
2.0\times 10^{-6}$. The second glitch, separated from the first by about 3.2
years, has a fractional amplitude of $\sim 1.2 \times 10^{-6}$, making it also
a member of the giant glitch class. 

The longer timespan of the present observations following the first glitch
show that the post-glitch relaxation is best described by an exponential
recovery with a timescale of several hundred
days. It is likely that period irregularities were primarily responsible for
the apparent quicker relaxation found by Lyne et al. (1996). There is also
evidence for a similar long-timescale relaxation following the second
glitch. A TEMPO fit across both glitches yielded the parameters given
in Table~\ref{tb:glparam}, showing that, for both glitches, about 20 per cent
of the glitch decayed. For the first glitch, the fitted timescale was 800
days. No fit of the timescale to the second glitch was possible because of
the shorter data span; an assumed decay time of 600 days gave a minimum in the
post-fit residuals. 

\subsection{PSR J1803$-$2137, PSR B1800$-$21}
PSR J1803$-$2137 is a young pulsar ($\tau_c \sim 16,000$ y) discovered by
Clifton \& Lyne (1986)\nocite{cl86}. Lyne \& Shemar (1996) observed a large
glitch ($\Delta\nu_g/\nu \sim 4.1 \times 10^{-6}$) in 1990 December which
showed an exponential recovery with timescale of $\sim 150$ d, together
with an apparently linear decay in $\dot\nu$ indicating decay from an
earlier unobserved glitch. 

\begin{figure} 
\centerline{\psfig{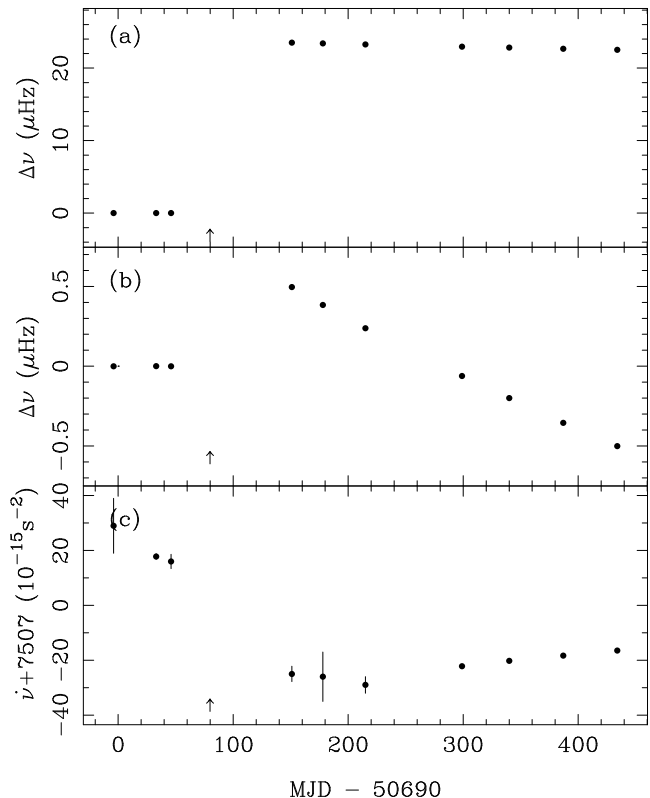}} 
\caption{The 1997 December glitch of PSR J1803$-$2137: variations of (a) frequency residual
$\Delta\nu$ relative to the pre-glitch solution, (b) an expanded plot of
$\Delta\nu$ where the mean residual after the glitch has been
removed, and (c) the variations of $\dot\nu$. }
\label{fg:1803_nu}
\end{figure}

Our observations of this pulsar are from 1997 August to 1998
December. Fig.~\ref{fg:1803_nu} shows that another large glitch of magnitude
$\Delta\nu_g/\nu \sim 3.2 \times 10^{-6}$ occurred in 1997 November. Again,
there the signature of an exponential decay following the
glitch is seen. There is, however, no evidence of decay in $\dot\nu$ preceding the
glitch (Table~\ref{tb:pparam}), although the data span is rather short.
Table~\ref{tb:glparam} gives the extrapolated and fitted parameters for the
glitch. Fitting of a single exponential decay gives a relatively good fit to
the post-glitch data with $Q\sim 0.13$ and $\tau_d \sim 640$ d and an rms
residual of 1690 $\mu$s. However, there is clear evidence in the residuals
immediately after the glitch for a shorter-term decay. The glitch parameters
in Table~\ref{tb:glparam} are from a simultaneous fit of two exponential
decays, one with a short decay time and the other representing the
longer-term decay. The final fit is extremely good, with a final rms
residual of only 334 $\mu$s and the residuals dominated by random
noise. Fitting of the short-term decay resulted in an increase in the
estimates for $Q$ and $\tau_d$ for the longer-term decay by about 25 per
cent as shown in Table~\ref{tb:glparam}. 

\section{Discussion}
In this paper we have presented timing observations of 40 pulsars over
various intervals up to a maximum of 8.9 years and analysed these data for
improved astrometric and pulse frequency parameters and for glitch
activity. In total, 30 glitches were detected in 11 pulsars, including the
largest known glitch, with $\Delta\nu_g/\nu \sim 6.5 \times 10^{-6}$ in PSR
J1614$-$5047. Twelve glitches were detected in the period of PSR
J1341$-$6220 in a data span of 8.2 years, making this the most frequently
glitching pulsar known. Evidence was found in PSR J1614$-$5047 for a new class of
irregularity in which the pulse frequency increases markedly over a few-day
interval. There appears to be an accompanying decrease in the magnitude of
the frequency derivative, implying a corresponding decrease in braking
torque.

\begin{figure} 
\centerline{\psfig{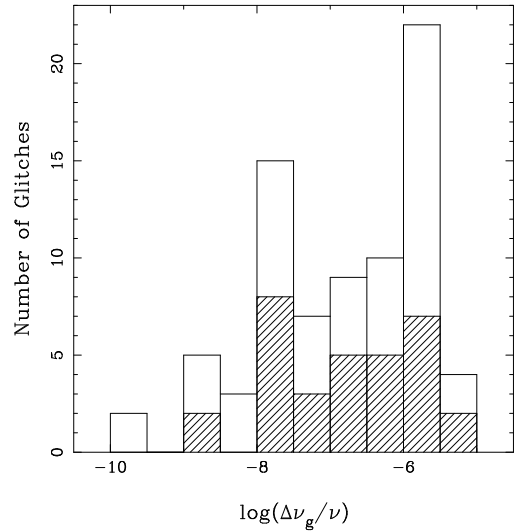}} 
\caption{Histogram of fractional glitch amplitudes $\Delta\nu_g/\nu$. Values
from the present study are cross-hatched. }
\label{fg:glt_hist}
\end{figure}

Table~\ref{tb:all_glt} lists all known glitches, 76 in total, giving the
fractional glitch amplitude $\Delta\nu_g/\nu$ and, in parentheses, the
approximate MJD of the glitch.  Fig.~\ref{fg:glt_hist} shows a histogram of
fractional glitch amplitudes.  Largely because of the Vela pulsar, the most
common glitches are large, with $\Delta\nu_g/\nu$ in the range $1 - 3 \times
10^{-6}$. In most other pulsars, smaller glitches with fractional amplitude
down to $10^{-8}$ are more common. Glitches with size smaller than $10^{-9}$
are difficult to identify, especially in noisy pulsars, and the sample is
certainly incomplete at this level. However, there does appear to be a
reduced rate of occurence of glitches with $\Delta\nu_g/\nu < 10^{-8}$.

\begin{table*}
\begin{minipage}{150mm}
\caption{Summary of all known glitches}
\begin{tabular}{lllll}
  ~~~PSR J & ~~PSR B & $N_g$ & ~~~$\Delta\nu_g/\nu$ & References \\ 
  &  &  & ~~~~($10^{-9}$) & \\ \\
0358+5413  & 0355+54 & 2 & 5.6(46079), 4368(46470) & 1 \\ 
0528+2200  & 0525+21 & 2 & 1.2(42057), 0.3(43810) & 2,3 \\ 
0534+2200  & 0531+21 & 5 & 9.5(40494), 37.2(42447), 9.2(46664), 85.0(47767), 4.7(48945) & 4,5 \\ 
0835$-$4510  & 0833$-$45 & 14 & 2336(40280), 2045(41192), 12(41312),
  1985(42683), 3060(43690), 1137(44889), & 6,7,8,9,10 \\
  & & & 2049(45192), 1601(46257), 1805(47520), 2715(48457), 5.6(48550), 861(49559), \\
  & & & 198(49591), 2150(50369) \\ 
1048$-$5832   & 1046-58   & 3  & 19(48944), 3000(49034), 769(50788) & 10  \\ 
1105$-$6107   &           & 2  & 279.7(50417), 2.1(50610) & 10 \\ 
1123$-$6259   &           & 1  & 749.31(49705)  & 10\\ 
1328$-$4357   & 1325$-$43 & 1  & 116(43590) & 11  \\ 
1341$-$6220   & 1338$-$62 & 12 & 1509(47989), 22.5(48453), 996(48645), 13.2(49134), 
  146(49363), 37(49523), &  3, 10, 12  \\ 
  & & & 15(49766), 31(49904), 1648(50008), 29.9(50321), 23.4(50528), 707.5(50683) \\
1509+5531    & 1508+55   & 1 & 0.22(41675) &  13 \\ 
1539$-$5626  & 1535$-$56 & 1 & 2790.4(48165) & 3, 14 \\ 
1614$-$5047  & 1610$-$50 & 1 & 6460(49802) &  10 \\ 
1644$-$4559  & 1641$-$45 & 3  & 191(43390), 803.9(46453), 1.9(47589) 
  & 15, 16, 17 \\ 
1709$-$4428  & 1706$-$44 & 1  & 2028(48778) & 3, 10, 14 \\ 
1730$-$3350  & 1727$-$33 & 1  & 3070(47990) & 3, 14 \\
1731$-$4744  & 1727$-$47 & 2  & 139.2(49387), 3.1(50703) & 10 \\ 
1739$-$2903  & 1736$-$29 & 1  & 2.9(46956) & 3, 14 \\ 
1740$-$3015  & 1737$-$30 & 9  & 427(46991), 31(47281), 35(47458), 601.9(47670)
  642(48186), 48(48218), & 3 \\ 
 & & & 15.7(48431), 10.0(49046), 169.6(49239) \\
1801$-$2304  & 1758$-$23 & 6  & 217(46907), 231.9(47855), 351(48454), 60.8(49701),
  17.0(50050), 87(50412) &  3, 10, 18\\ 
1801$-$2451  & 1757$-$24 & 2 &  1988(49476), 1248(50651) & 10, 19 \\ 
1803$-$2137  & 1800$-$21 & 2  & 4073(48245), 3185(50765) & 3, 10 \\ 
1826$-$1334  & 1823$-$13 & 2 & 2718(46507), 3049(49014)  & 3 \\
1833$-$0827  & 1830$-$08 & 1 & 1865.9(48041)  &  3  \\
1901+0716    & 1859+07   & 1 & 30(46859)   & 3 \\
2225+6535    & 2224+65   & 1 & 1707(43072) & 3, 20 \\  \\
\end{tabular}
\label{tb:all_glt}
1.~Lyne (1987)\nocite{lyn87} 2.~Downs (1982)\nocite{dow82} 3.~Shemar \& Lyne
   (1996)\nocite{sl96} 4.~Lohsen (1975)\nocite{loh75} 5.~Lyne et
al. (1993)\nocite{lps93} 6.~Cordes et al. (1988)\nocite{cdk88} 7.~McCulloch
et al. (1987)\nocite{mkhr87} 8.~McCulloch et al. (1990)\nocite{mhmk90}
9.~Flanagan (1995a)\nocite{fla95a}    10.~This paper   11.~Newton et al. (1981)\nocite{nmc81}  12.~Kaspi et al. (1992)\nocite{kmj+92}   13.~Manchester et al.(1974)\nocite{mt74}  14.~Johnston et al. (1995)\nocite{jml+95}  15.~Manchester et al.(1978)\nocite{mngh78}   16.~Flanagan (1993)\nocite{fla93}   17.~Flanagan (1995b)\nocite{fla95}   18.~Kaspi et al. (1993)\nocite{klm+93}  19.~Lyne et al. (1996a)\nocite{lkb+96} 20.~Lyne (1996)\nocite{lyn96c} 
\end{minipage}
\end{table*}

In most models for the glitch phenomenon, the sudden spin-up is triggered by
the release of stress built up as a result of the steady spin down of the
pulsar. For the original star-quake model \cite{bppr69}, the equilibrium
shape of the star becomes less oblate as the star spins down. At some point,
the crust cracks and relaxes to (or toward) the new equilibrium shape,
reducing its moment of inertia and hence spinning it up. However, this model
fails to account for frequent giant glitches, as seen, for example, in the
Vela pulsar, as the rate of change of oblateness is too slow. In models
based on unpinning of internal superfluid vortices \cite{acp89,rzc98}, the
stresses on the pinned vortices build up as the crust slows down until
finally some fraction of them unpin and then repin at a larger radius,
resulting in a transfer of angular momentum to the crust.

\begin{figure*}
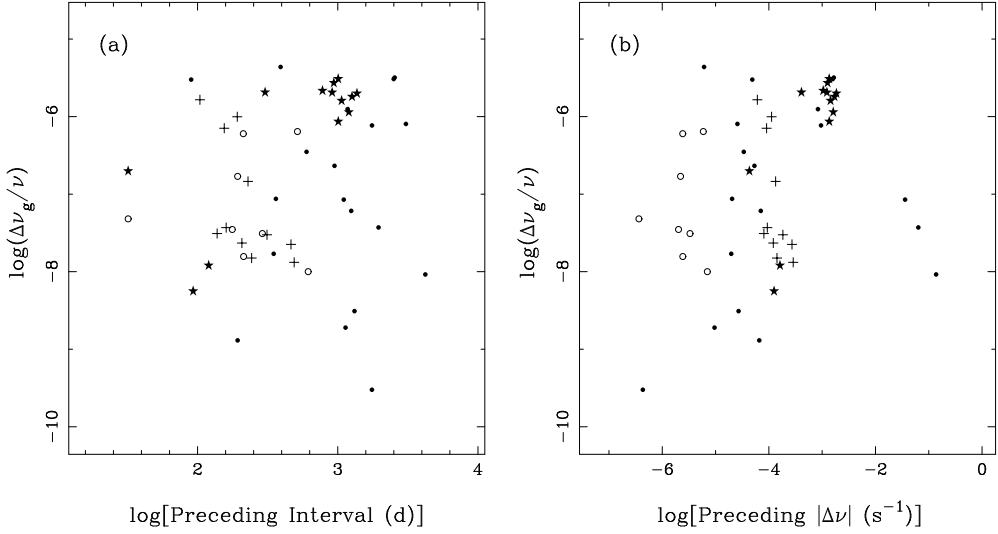
 
\begin{tabular}{cc}
\psfig{file=glt_int_prec.ps,height=70mm} &
\psfig{file=glt_dnu_prec.ps,height=70mm}
\end{tabular}
\caption{(a) Fractional glitch amplitude $\Delta\nu_g/\nu$ versus length of the
preceding inter-glitch interval. (b) Fractional glitch amplitude $\Delta\nu_g/\nu$ versus accumulated
spin-down frequency during the preceding inter-glitch interval. In both
plots, points for the Vela pulsar (PSR J0835$-$4510) are marked with a
$\star$, those for PSR J1341$-$6220 are marked with +, and those for PSR
J1740$-$3015 (PSR B1737$-$30) are marked with $\circ$.}
\label{fg:int_prec}
\end{figure*}

In these models, one expects some relation between the size of the
glitch and the length of time that the pulsar has been slowing down since
the previous glitch or the integrated change in spin rate since the last
glitch. Fig.~\ref{fg:int_prec}(a) shows glitch fractional amplitudes plotted
against the length of the preceding inter-glitch interval ($\Delta t_g$),
and in Fig.~\ref{fg:int_prec}(b) against the change in spin frequency since
the previous glitch ($|\dot\nu|\Delta t_g$). These figures show that,
contrary to expectations, there is no general relation between fractional
glitch amplitude and either the time since the previous glitch or the total
change in spin frequency since the previous glitch. For the Vela pulsar
(marked with a $\star$ in the Figure), the smaller glitches do tend to have
shorter preceding intervals and all but one of the giant glitches have
preceding inter-glitch intervals of about 1000 days. However, for PSR
J1341$-$6220 there is if anything an inverse relationship, with larger
glitches occuring after shorter intervals. For PSR J1740$-$3015 there is no
relationship between glitch size and preceding
interval. Fig.~\ref{fg:int_prec}(b) shows that similar relationships for
individual pulsars hold when glitch size is plotted against accumulated
spin-down since the last glitch. The three points to the right on this plot
are for the Crab pulsar; these too show an inverse relationship.

\begin{figure*}
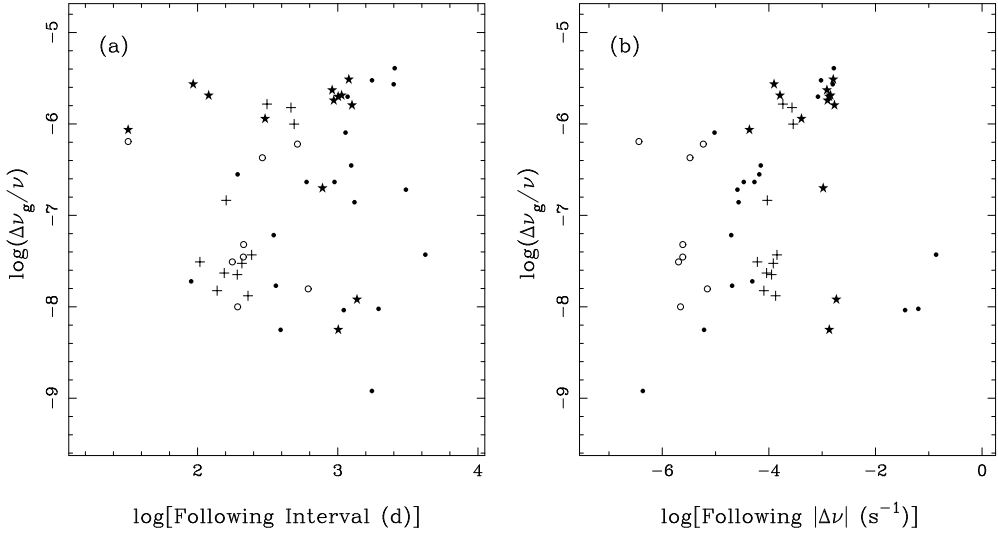
 
\begin{tabular}{cc}
\psfig{file=glt_int_foll.ps,height=70mm} &
\psfig{file=glt_dnu_foll.ps,height=70mm}
\end{tabular}
\caption{(a) Fractional glitch amplitude $\Delta\nu_g/\nu$ versus length of
the following inter-glitch interval. (b) Fractional glitch amplitude
$\Delta\nu_g/\nu$ versus accumulated spin-down frequency during the
following inter-glitch interval. In both plots, points for the Vela pulsar
(PSR J0835$-$4510) are marked with a $\star$, those for PSR J1341$-$6220 are
marked with +, and those for PSR J1740$-$3015 (PSR B1737$-$30) are marked
with $\circ$.}
\label{fg:int_foll}
\end{figure*}

Alternatively, if a small glitch results from a release of only part of the
built-up stress, one might expect another glitch to occur soon after, when
the breaking strain is again reached. Conversely, if a large glitch releases
all or most of the stress, a long time would be required for it to build up
again. This suggests a correlation
between glitch size and time to the {\it next}
glitch. Fig.~\ref{fg:int_foll} shows glitch size plotted against duration of
the following inter-glitch interval and accumulated spin-down in that
period. No such correlation is observed for the Vela pulsar, where the small
glitches are followed by relatively long intervals, but a weak positive
correlation is seen for PSR J1341$-$6220 and PSR J1740$-$3015. Excepting the
Crab pulsar, other pulsars (marked with a dot) have a good correlation
between glitch size and accumulated spin-down frequency following the
glitch. 

These results suggest that the triggering of glitches is a local phenomenon, not
dependent on global stresses. This lends some support to the ideas of
Ruderman et al. (1998) in which migration of magnetic flux tubes determines
the stresses on the pinned vortices. This model is also supported by the
observations of the ``slow'' glitches and subsequent decrease in slow-down
rate in PSR J1614$-$5047. 

\begin{figure*}
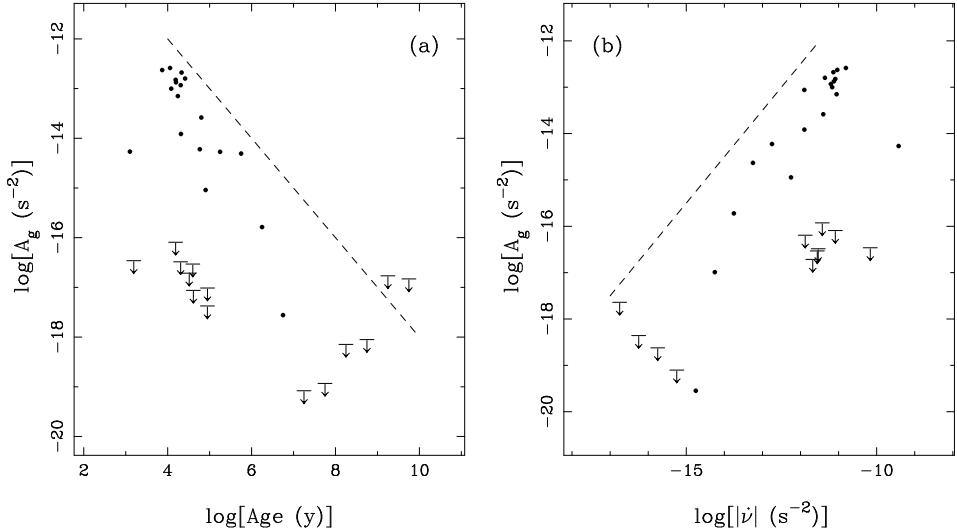
 
\begin{tabular}{cc}
\psfig{file=ag_age.ps,height=70mm} &
\psfig{file=ag_dnu.ps,height=70mm}
\end{tabular}
\caption{(a) Plot of glitch activity parameter, $A_g$, versus characteristic
age $\tau_c$. Pulsars with $\tau_c < 10^5$ yr are plotted
individually, whereas those with greater $\tau_c > 10^5$ yr are mean values in
half-decade bins. Where no glitch was observed, the upper limit is based on
a single glitch of fractional amplitude $10^{-9}$ over the total data span,
summed over all pulsars in the case of binned data. Upper limits are higher
for the very old pulsars simply because there are fewer of them known and
the total accumulated data span is less. The dashed line has a slope of
$-1.0$. (b) A similar plot with the absolute value of the spindown rate
$\dot\nu$ as abscissa. Points with $|\dot\nu| > 10^{-12}$ are plotted
separately while those with $|\dot\nu|$ less than this value are binned.}
\label{fg:ag}
\end{figure*}

McKenna \& Lyne (1990)\nocite{ml90} introduced the glitch activity
parameter, $A_g$, defined to be the accumulated pulse frequency change
$\Delta\nu_g$ due to glitches divided by the observation data span. It
therefore has the same units as $\dot\nu$ and represents the portion of
$\dot\nu$ which is overcome by glitches. This is typically small; the
largest known value of $A_g$, for the Vela pulsar, is $\sim 2.5 \times
10^{-13}$ s$^{-2}$, only about 2 per cent of its spin-down rate $\dot\nu
\sim 1.6 \times 10^{-11}$ s$^{-2}$. Fig.~\ref{fg:ag}(a) shows activity
parameter plotted against characteristic age, $\tau_c$ for most of the
pulsars with extensive timing data. As noted by McKenna \& Lyne (1990),
there is a clear peak in activity for pulsars with ages between 2,000 and
20,000 years. 

However, there is a substantial group of young pulsars with low glitch
activity; these so far have had no observed glitch and are represented by
upper limits in Fig.~\ref{fg:ag}(a) corresponding to a single glitch of
fractional size $10^{-9}$. These eight pulsars have a mean data span of more
than six years and so, at the least, they are infrequent glitchers. A single
glitch of amplitude $\sim 10^{-6}$ would raise them into the same region as
the other young pulsars. The dashed line corresponds to a uniform
accumulated $\dot\nu$ over the characteristic spin-down time. This is
approximately true for pulsars with age of less than or about $10^6$ years,
but older pulsars clearly have less glitch activity than predicted by this
rule.

In Fig.~\ref{fg:ag}(b) the activity parameter is plotted against the absolute
value of $\dot\nu$. Some theoretical models for glitches (e.g. Ruderman, Zhu
\& Chen, 1998)\nocite{rzc98} predict that the ratio of the effective spin-up
rate due to glitches to the spin-down rate is proportional to the ratio of
the moment of inertia of the crustal superfluid to the total moment of
inertia of the neutron star. A constant ratio corresponds to a line of slope
$-1$ in Fig.~\ref{fg:ag}(b). For the younger pulsars, except PSR B0531+21
(Crab) and PSR B1509$-$58, the points correspond to a fraction of 1 -- 2 per
cent (cf. Lyne et al. 1999)\nocite{lss99}. Pulsars with 
spin-down rates less than about $10^{-14}$ s$^{-1}$ have a significantly
smaller fraction of their spindown recovered by glitches.

\begin{figure*}
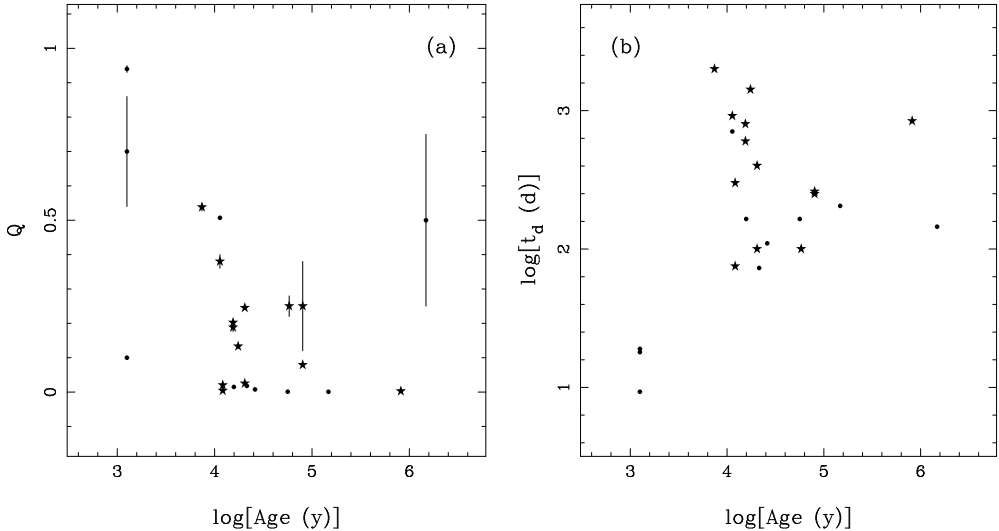
 
\begin{tabular}{cc}
\psfig{file=q_age.ps,height=70mm} &
\psfig{file=td_age.ps,height=70mm}
\end{tabular}
\caption{(a) Plot of $Q$, the fraction of $\Delta\nu_g$ which decays, versus
characteristic age. (b) Plot of glitch decay timescale versus characteristic
age. Results from this work are marked with a $\star$.}
\label{fg:q_td}
\end{figure*}

When a significant post-glitch decay is observed, it is generally well
described by the exponential relation given in Equation~\ref{eq:glmodel}. In
Fig.~\ref{fg:q_td}(a) the fraction of the glitch which decays, $Q =
\Delta\nu_d/\Delta\nu_g$ is plotted against characteristic age. With one
exception, PSR B0525+21, old pulsars tend to have low values of $Q$. Younger
pulsars can have larger $Q$, but many glitches in young pulsars have little
or no decay. 

Fig.~\ref{fg:q_td}(b) is a plot of glitch decay timescale  versus pulsar
characteristic age. This shows that, apart from the Crab pulsar (the three
points in the lower left of the figure) in which the decay timescale is very
short, there is no relation between decay timescale and pulsar
age. We emphasise that this result applies to the longest decay timescale
present. This may indicate the unreliability of characteristic age as an
indicator of true age and hence internal temperature, or it may indicate
that decay time, at least for pulsars with age greater than a few thousand
years, is dependent on other properties, for example, the core magnetic field.

These observations have demonstrated the great diversity of glitch
properties. Glitch activity is clearly greatest in pulsars with ages of
between a few times $10^3$ and $10^5$ years. The Crab pulsar has distinctly
different glitch properties from those of the middle-aged pulsars, and other
pulsars of similar age show no glitches at all. Apart from these clear
trends, glitch properties vary greatly, both for successive glitches from a
given pulsar and across different pulsars, with few systematic trends. These
properties suggest that the glitch phenomenon, both the event and the
following response,  depend on quasi-random
processes occurring in the pulsar, rather than global properties such as
slow-down rate or age. Crustal processes driven by magnetic field evolution
as proposed by Ruderman et al. (1998) seem more consistent with this than
the vortex creep models of Alpar et al. (1989).

\section*{ACKNOWLEDGEMENTS}
We thank the many colleagues who have helped with observing and software
development over the course of this project, and the staff of Parkes
Observatory for their always cheerful assistance. In particular, we thank
John Sarkissian for doing much of the observing during 1998, and Matthew
Britton, Stuart Anderson and John Yamasaki for help with correlator hardware
and software development. NW thanks the National Nature Science Foundation
(NNSF) of China for their support of this work and VMK acknowledges support
from an Alfred P. Sloan Research Fellowship. The Parkes radio telescope
is part of the Australia Telescope which is funded by the Commonwealth of
Australia for operation as a National Facility managed by CSIRO.

\bibliographystyle{mn}
\bibliography{modrefs,psrrefs,crossrefs}

\end{document}